\documentclass[a4paper,11pt]{article}
\usepackage{jheppub}
\usepackage{amsmath}
\usepackage{amsthm}
\usepackage{amsfonts}
\usepackage{amssymb}
\usepackage{textcomp}
\usepackage{graphicx}
\usepackage{bm}
\usepackage{color}
\usepackage{verbatim}
\usepackage{subcaption}
\usepackage{graphicx,color}
\usepackage{epsfig}

\usepackage{amsfonts,amsmath,amssymb}

\newcommand{\rar}{\rightarrow}

\theoremstyle{definition}
\newtheorem{theorem}{Proposition}

\begin{document}

 \title{Commensurate lock-in in holographic non-homogeneous lattices}
 
 \author[a]{Tomas Andrade}
  
 \author[b,c]{Alexander Krikun\footnote{https://orcid.org/0000-0001-8789-8703}}

 \affiliation[a]{Rudolf Peierls Centre for Theoretical Physics \\ University of Oxford, 1 Keble Road, Oxford OX1 3NP, UK} 

 \affiliation[b]{Instituut-Lorentz, Universiteit Leiden \\ P.O. Box 9506, 2300 RA Leiden, The Netherlands} 
 \affiliation[c]{Institute for Theoretical and Experimental Physics (ITEP)\footnote{On leave from}, \\ B. Cheryomushkinskaya 25, 117218 Moscow, Russia }

\emailAdd{tomas.andrade@physics.ox.ac.uk}
\emailAdd{krikun@lorentz.leidenuniv.nl}

\abstract{
We consider the spontaneous formation of striped structures in a holographic model which possesses explicit translational 
symmetry breaking, dual to an ionic lattice with spatially modulated chemical potential. 
We focus on the perturbative study of the marginal modes which drive the transition to a phase exhibiting spontaneous stripes. We study the wave-vectors of the instabilities with largest critical temperature in a wide range of backgrounds characterized by the period and the amplitude of the chemical potential modulation.

We report the first holographic observation of the commensurate lock-in between the spontaneous stripes and the underlying ionic lattice, 
which takes place when the amplitude of the lattice is large enough. We also observe an incommensurate regime in which the amplitude of the lattice is finite, but the preferred stripe wave-vector is different from that of the lattice. 

}

\maketitle

\section{Introduction}

The study of momentum relaxation in holography has recently received a great deal of attention. One of the main reasons behind this interest is that, by breaking translational invariance, the lattice provides a mechanism to dissipate momentum so the (thermo)electric resistivity acquires a finite value thus making the holographic descriptions more realistic. There are by now many holographic models which consider lattices and study interesting metallic, insulating and superconducting phases as well as the transitions between them 
\cite{Donos:2012js, Donos:2013eha, Andrade:2013gsa, Donos:2014oha, Donos:2014uba, Gouteraux:2014hca, Taylor:2014tka, Horowitz:2012ky, Horowitz:2013jaa, Donos:2014yya, Rangamani:2015hka, Erdmenger:2015qqa, Andrade:2014xca, Kim:2015dna, Ling:2014laa, Ling:2014saa, Baggioli:2015zoa, Baggioli:2015dwa}. 

Depending on how translational invariance is broken, one classifies lattices in holographic models into {\it non-homogeneous} and  {\it homogeneous} ones. In the former case, the sources are periodic along a generic spatial direction of the boundary. 
As a result of this genuine spatial dependence, the equations of motion that govern these stationary solutions are 
partial differential equations (PDE's) in the holographic and boundary directions. The first non-linear such example was considered in 
\cite{Horowitz:2012ky}, and many further developments have been made, 
see e.g. \cite{Horowitz:2012gs, Donos:2014yya, Rangamani:2015hka, Liu:2012tr, Ling:2013aya}. In the homogeneous case, on the other hand, 
translational invariance is broken along a certain global symmetry direction which compensates for it, simplifying the spatial 
dependence of the equations of motion and reducing them to ordinary differential equations (ODE's). Examples of this kind include 
\cite{Donos:2012js, Donos:2013eha, Andrade:2013gsa}\footnote{Another mechanism of momentum dissipation 
that has been studied in holography is to break diffeomorphism invariance by introducing a mass for the graviton 
in four bulk dimensions by considering the theory of Massive Gravity \cite{Vegh:2013sk}. The connection of
this mechanism and the introduction of a perturbative lattice was later understood in \cite{Blake:2013owa}.}.
Similar setups have been later used to describe holographic metal-insulator transitions\footnote{Most insulators in holography do not have a gap, but are defined as $\lim_{T\rar0}\sigma_{DC}(T) = 0$} and ``strange metals'' \cite{Donos:2013eha, Donos:2012js, Ling:2015dma, Hartnoll:2015sea, Lucas:2014sba}. In all these (non)-homogeneous examples, the system relaxes momentum and the 
DC conductivities are finite. 

In addition to relaxing momentum, the real world ionic lattice introduces a length scale which plays an important role in commensurability 
effects\footnote{Here we will be concerned only with monochromatic lattices, for which the characteristic scale is simply the 
period of the background structure.}. One such effect is what characterizes a Mott 
insulator. Briefly, this state emerges when a substrate ionic lattice becomes commensurate with a spontaneously formed electronic Wigner crystal, localizing the charge carriers which in turn yields an insulating state at half-filing of the otherwise metallic electronic bands. Other examples include anti-ferromagnet, where the spin density wave is commensurate with the lattice, or ferroelectric, where the spontaneous wave of displacements plays the same role.
Even more interesting are the phase transitions between the commensurate and incommensurate phases, which exhibit salient phenomena including incommensurate spin density waves, discommensuraton lattices etc. The holographic description of these commensurability effects is the main subject of the present study.

In order to exhibit commensurability, apart from the background ionic lattice, the model under consideration should include an additional spatially modulated structure, dual to an electronic lattice. The holographic models describing such spontaneous structures on the translationally invariant backgrounds are well developed.   
Among the models of this type we mention the spontaneous formation of helices \cite{Nakamura:2009tf, Ooguri:2010kt, 
Donos:2012wi}, charge density waves \cite{Donos:2013gda}, helical superconductors \cite{Donos:2011ff} and various types of striped orders \cite{Donos:2011bh, Donos:2011pn, Donos:2011qt, Rozali:2012es, Donos:2013woa, Withers:2013kva, Withers:2013loa, Jokela:2014dba, Withers:2014sja, Krikun:2015tga, Erdmenger:2013zaa, Cremonini:2016rbd, Ling:2014saa}. 
In all these models below a certain temperature the translationally invariant Reissner-Nordstr\"om (RN) solution exhibits 
near horizon instabilities in dynamical modes with specific nonzero momenta signalling the formation of spontaneous stripes.
In the presence of an UV-sourced ionic lattice that modifies the near-horizon geometry
one expects that the explicit breaking of translational invariance would affect the characteristic momenta (and corresponding critical temperature) of the stripe, in a way which is correlated to that of the substrate lattice. 
This issue was investigated in \cite{Andrade:2015iyf} in the context of holographic homogeneous lattices\footnote{See also a relevant study of stripes on homogeneous background \cite{Jokela:2016xuy}, where somewhat different questions were addressed.}. It was found there that the instabilities signalling the formation of spontaneous stripes are indeed only sensitive to the near horizon 
geometry of the black hole solutions, but that their specifics cannot be related to the length scale of the background 
lattice, at least not in a simple way which would reveal commensurability. 
The absence of the effect can perhaps be explained by recalling that, because of the particular structure 
of holographic homogeneous lattices, the equations that govern the relevant degrees of freedom are ODE's 
so the traces of the spatial dependence are somehow lost. 

Motivated by this, we undertake the study of commensurability in non-homogeneous lattices. 
In this setup, the equations that describe the physics are PDE's so it seems more reasonable that the interaction
of commensurate structures will be manifest. 
We choose to work with the ionic lattice resulting from introducing 
a spatially modulated chemical potential in the Einstein-Maxwell system in four bulk dimensions \cite{Horowitz:2012gs}. The choice of this model is mostly motivated by its simplicity (only the metric and a gauge field are present in the bulk) and consequent generality.
This lattice is also known to be irrelevant in the IR at zero temperature, so the black hole corresponds to a metallic state in the dual theory 
\cite{Donos:2014yya}. As we will see, however, at finite temperature the near horizon 
spatial modulation is finite and one should expect the marginal modes to be affected by this modulation.

The specific mechanism of spontaneous stripe formation which we consider is that of \cite{Donos:2011bh}. It consists of a pseudo-scalar field $\psi$ that couples to the Maxwell field via the term 
$\sim \psi F \wedge F$ in the Lagrangian. In \cite{Donos:2011bh} the marginal modes around RN were obtained via a linearized 
analysis and it was shown that the temperatures where each individual mode with given wavenumber becomes unstable arrange themselves as a characteristic ``bell''-shaped instability curve, with the maximum critical temperature exhibited by a distinct mode with finite momentum depending on the strength of the $\psi F \wedge F$ term.

Combining features of these two models, we will study the formation of stripes on top of an ionic lattice by considering the 
Einstein-Maxwell-pseudo-scalar system of \cite{Donos:2011bh} and choosing boundary conditions corresponding 
to a spatially modulated chemical potential as in \cite{Horowitz:2012gs}. 
More precisely, we will study the spatially modulated instabilities of the holographic backgrounds with finite amplitude ionic lattice, trace the corresponding instability curves and look for the parameters of the marginal mode with highest critical temperature for a given amplitude and period of the lattice. Above a certain amplitude, we will find that the preferred wave-vector 
of the stripe deviates from its value in the absence of a lattice and prefers to coincide with multiples of the lattice wave-vectors, showing 
the commensurate lock-in between two structures. At smaller amplitudes we encounter an incommensurate regime in which the preferred wave-vector ranges between the RN value and the one dictated by the lattice.

This paper is organized as follows. To begin with, we review the basic aspects of the physics of commensurate lock-in by considering 
a simple mechanical toy model in Sec.\,\ref{sec:physics_of_commensurate_lock}, which will help us in the analysis of the 
phenomena we observe in holography. We introduce the details of our holographic model in Sec.\,\ref{sec:the_model}. We perform the study of modulated marginal modes in Sec.\,\ref{sec:marginal_modes} and discuss the obtained results is Sec.\,\ref{sec:discussion}. In Appendix \ref{app:precision} we comment on some precision issues arising in the numerical linear stability analysis.

Throughout the paper we will use the term ``lattice'' to denote the explicit translational symmetry breaking structure, sourced in UV, and ``stripe'' for the spontaneously formed translational symmetry breaking structure arising due to the dynamical instabilities near the horizon. The wave-vectors and wavelengths (periods) of the lattice and stripe will be denoted as
\begin{align}
\label{equ:notation}
\mathrm{lattice:} \qquad &\lambda_k \equiv 2\pi / k, \\
\notag
\mathrm{stripes:} \qquad &\lambda_p \equiv 2\pi / p,
\end{align}
\noindent where $k$ and $p$ are the corresponding wave-vectors.

\section{Physics of commensurate lock-in} 
\label{sec:physics_of_commensurate_lock}

\begin{figure}[ht]
\centering
\begin{subfigure}{0.3\linewidth}
\includegraphics[width= \linewidth]{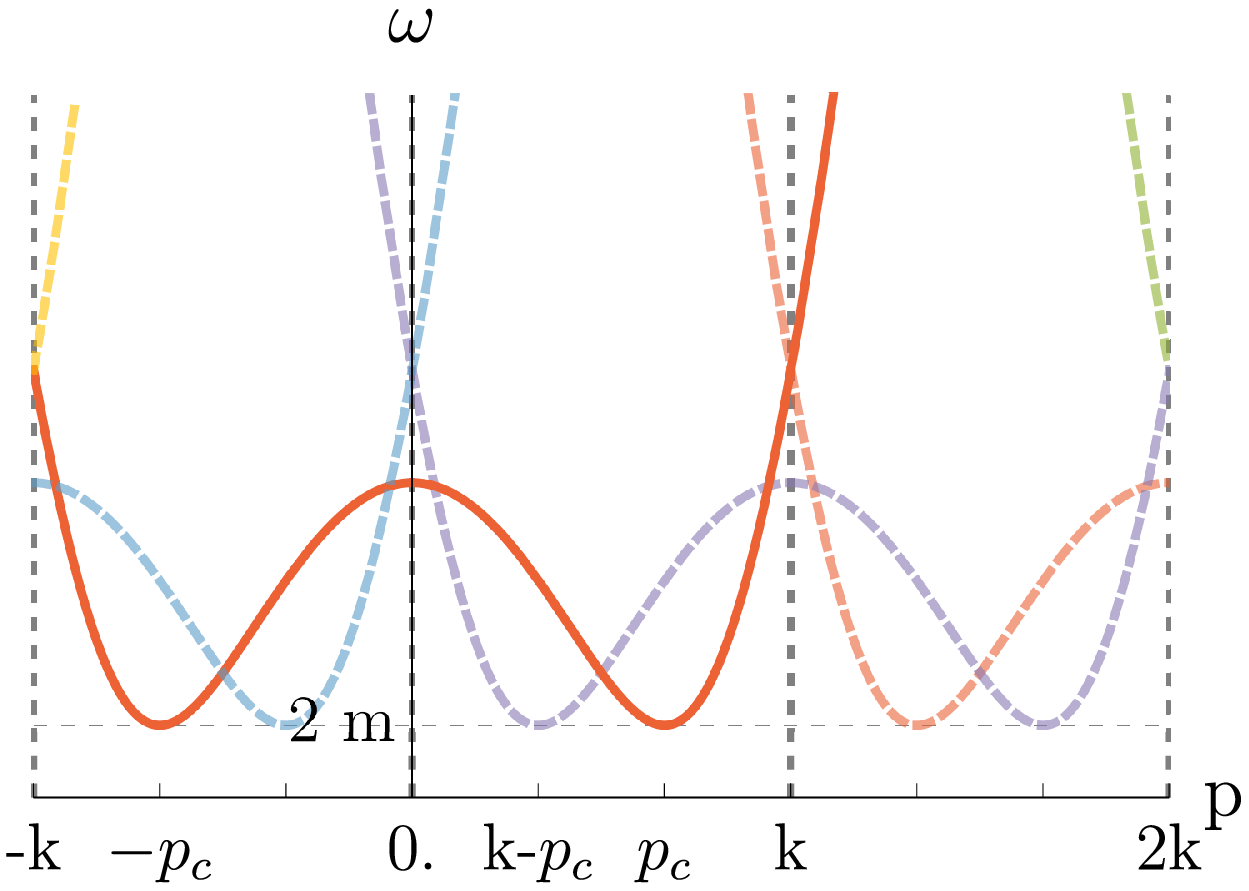}
\caption{\label{fig:double-well1} Generic lattice}
\end{subfigure} \quad
\begin{subfigure}{0.3\linewidth}
\includegraphics[width= \linewidth]{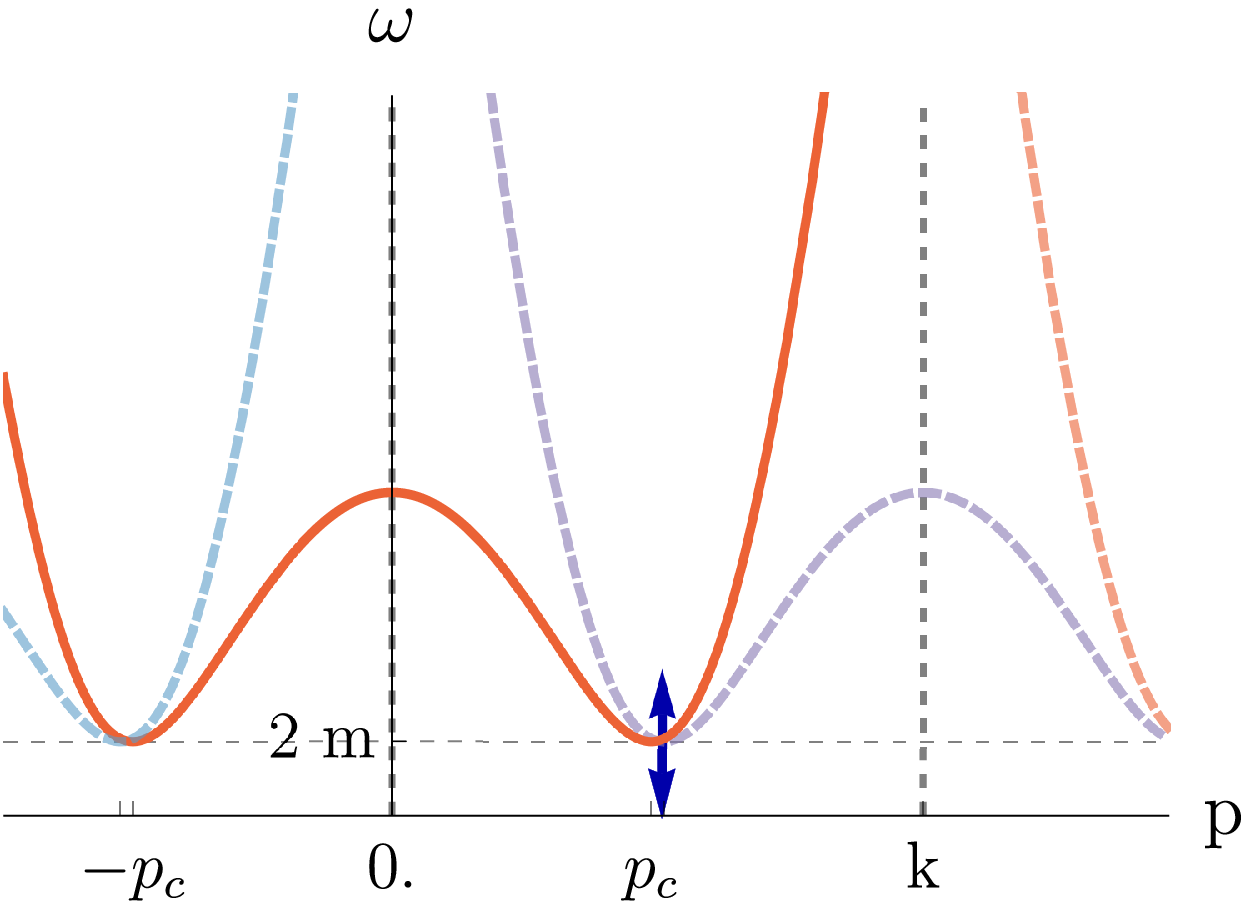}
\caption{\label{fig:double-well2} Commensurate lattice}
\end{subfigure} \quad
\begin{subfigure}{0.3\linewidth}
\includegraphics[width= \linewidth]{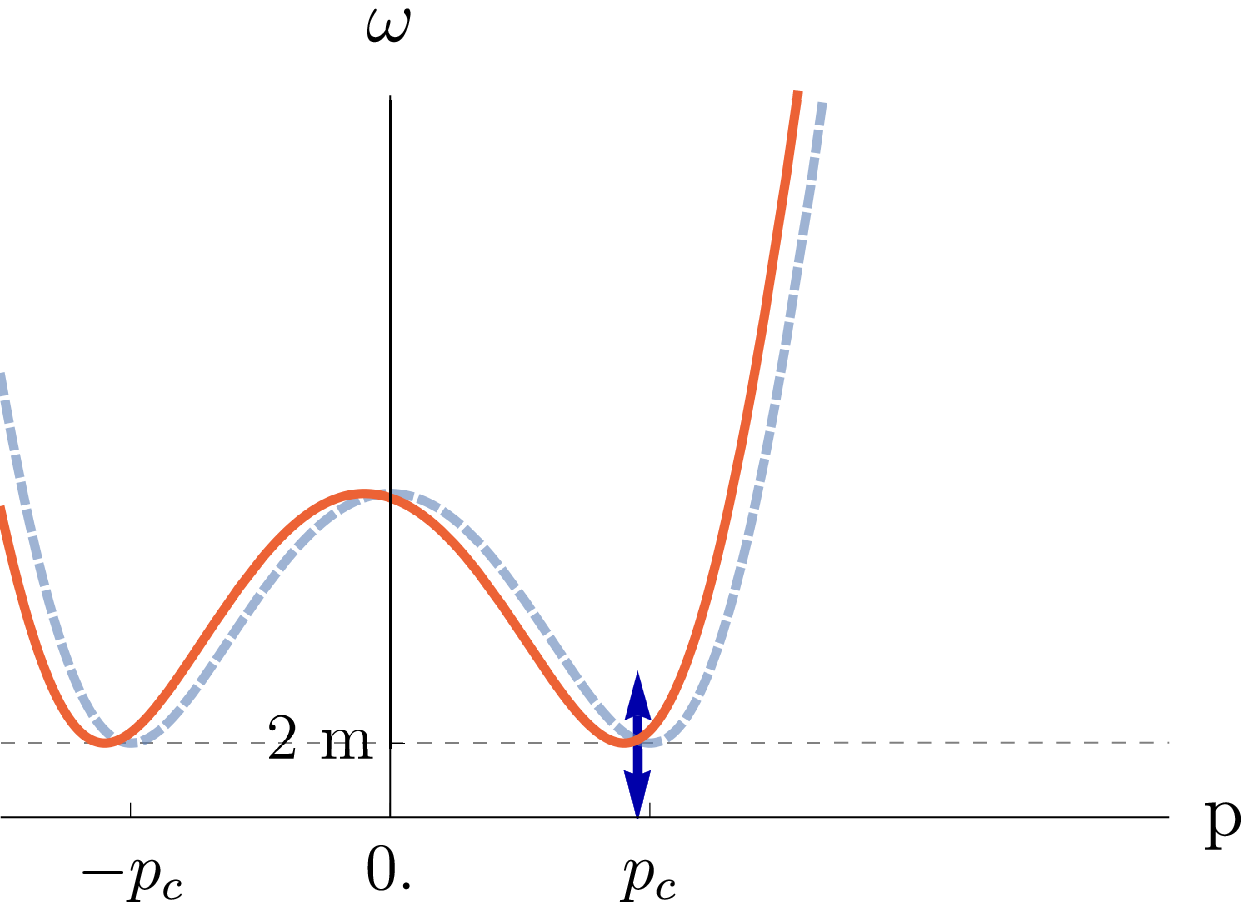}
\caption{\label{fig:double-well3} Long-wavelength lattice}
\end{subfigure} 
\caption{Dispersion relation in the ``double-well'' toy model on top of a periodic background lattice with different wave-vectors. The dashed curves represent the ``shifted'' dispersions $\omega_n(\tilde{p})=\sqrt{D_{\tilde{p}+nk}}$. The blue arrows show the eigenmode repulsion leading to decrease of the ground state energy in the commensurate and long-wavelength cases.}
\end{figure}

Before we proceed with the study of the holographic model, it is instructive to consider the qualitative features of the physics we are after by making use of a simple toy model as an example. The following considerations are quite standard in condensed matter theory and can be found in one's favorite handbook, see e.g. \cite{kittel1966introduction}, but we find it convenient to restate them here for consistency and reference purposes. The physics of spontaneous translation symmetry breaking can be captured by the mechanics of a particle with quartic ``double-well'' dispersion relation. 
The equation of motion reads
\begin{equation}
\label{eq:double-well_potential}
 D_p \phi(p) = \hat{\omega}^2 \phi(p), \qquad D_p \equiv \left(\frac{p^4 + p_0^4}{2 p_0^2} -  p^2   + 4 m^2 \right). 
\end{equation} 
It has two minima located at $p=\pm p_0$, hence the classical ground state breaks parity and translational symmetry spontaneously, see Fig.\,\ref{fig:double-well1}. The dispersion near the minimum assumes the form  
\begin{equation}
\label{equ:mass}
\hat{\omega}|_{p \rar p_0} = 2 m + \frac{1}{2 m} (p-p_0)^2+ O[(p-p_0)^3].
\end{equation}
The linear excitations are particles with mass $m$. The ground state energy defined as
\begin{equation}
\label{equ:ground_state_energy}
\varepsilon \equiv \mathrm{min}[\omega(p)] 
\end{equation}
equals $\varepsilon_0=2m$. From the thermodynamics of the second order phase transitions, which will be the focus of our discussion in the following Sections, given the Ginzburg-Landau free energy behaves near the critical temperature as $\mathcal{F}(\Psi) \sim \alpha' (T-T_c)|\Psi|^2$, we expect that the decrease of the ground state energy leads roughly to the increase of critical temperature. Hence we will be mostly interested in the lowest energy state in the remainder of this Section.

Now let us include the infinitesimal periodic potential in the model \eqref{eq:double-well_potential}. For simplicity we will consider one harmonic mode with wave-vector $k$: $V(x) = A \cos(k x) \phi(x)^2$. Note that $A$ has dimension 2 in mass units. In the presence of the 
modulated potential, the equation of motion acquires a mixing term
\begin{equation}
\label{eq:mode_coupling}
D_p \phi(p) + \frac{A}{2}\left[\phi(p+k) +  \phi(p-k)\right]  = \omega^2 \phi(p).
\end{equation}
Note that the explicit breaking of continuous translations down to discrete shifts with given period introduces a coupling between modes and has a dramatic effect on the spectrum. In fact, in the translationally invariant model, the spectrum consisted of decoupled modes $\phi(p)$ characterized by continuum parameter $p \in \mathbb{R}$. Once the periodic potential is included, the spectrum breaks down into decoupled orbits $\{\phi(\tilde{p}),\phi(\tilde{p}+k),\dots,\phi(\tilde{p}+ n k)\}, \ n \in \mathbb{Z}$ of the remaining discrete symmetry, characterized by continuum $\tilde{p}$ -- the Bloch pseudo-momentum -- which is defined inside the Brillouin zone (BZ) $\tilde{p} \in (0,k)$\footnote{The choice of the actual interval representing the BZ is arbitrary and our choice may differ from some standard definitions $(-k/2,k/2)$}.
It is also instructive to write down explicitly the position space field as a Bloch wave function
\begin{equation}
\label{eq:Bloch_function}
	\phi(x) = e^{i \tilde p x} \sum_{n} \phi(\tilde p + n k) e^{i n k x}. 
\end{equation}
Note that the values of $\tilde{p}$ outside of the BZ can be absorbed by redefining the coefficients
$\phi(\tilde p + n k) $.  
The coupled equations of motion can be written in the matrix notation
\begin{equation}
\label{eq:matrix_coupling}
\begin{pmatrix}
\ddots & \ddots & & 0 \\
\ddots & D_{\tilde{p}-k} & A/2 &  \\
 & A/2 & D_{\tilde{p}} & \ddots \\
 0 & & \ddots & \ddots
\end{pmatrix}
\begin{pmatrix}
\vdots \\ \phi(\tilde{p}-k) \\ \phi(\tilde{p}) \\ \vdots
\end{pmatrix}
=
\omega^2
\begin{pmatrix} 
\vdots \\ \phi(\tilde{p}-k) \\ \phi(\tilde{p}) \\ \vdots
\end{pmatrix}.
\end{equation}
Considered as an eigenvalue problem, this equation has an infinite discrete set of solutions $\omega = \omega_n(\tilde{p}), \, n \in \mathbb{Z}$, which in the case $A=0$ would simply be $\omega_n(\tilde{p}) = \sqrt{D_{\tilde{p}+n k}} = \hat{\omega}(\tilde{p} + n k)$, where $\hat{\omega}$ is defined by \eqref{eq:double-well_potential}. In this way the explicit breaking of translational symmetry itself (even in the limit of vanishing $A$) transforms a single valued function $\hat{\omega} (p)$ defined on the real axis $\mathbb{R}$ to the multi-valued function $\omega_n (\tilde{p})$ defined on the interval $[0,k)$. The dispersion relation of the initial model is effectively folded into the BZ, see Fig \ref{fig:double-well1}.
	
Now consider the effects of nonzero inter-mode coupling $A$. Generically, the off-diagonal terms in \eqref{eq:matrix_coupling} lead to only quadratic in $A$ corrections to the eigenvalues. But in the special case when two neighbouring diagonal terms coincide, i.e. $D_{\tilde{p}-k} = D_{\tilde{p}}$, the ``repulsion of eigenvalues'' happens already to first order in $A$. In this case we can focus on the two degenerate modes and diagonalize the corresponding $2\times2$ block in \eqref{eq:matrix_coupling} explicitly. We get that if $D_{\tilde{p}-k}=D_{\tilde{p}} = \hat{\omega}(\tilde{p})^2$ then the corresponding perturbed eigenvalues are $\omega_{-1}(\tilde{p})^2 = \hat{\omega}(\tilde{p})^2 + A/2$ and $\omega_0(\tilde{p})^2 = \hat{\omega}(\tilde{p})^2 - A/2$. In the more general case of finite amplitude of the potential and/or presence of the higher harmonics in its profile, the couplings between modes separated by $2k, 3k, \dots$ become relevant as well and the repulsion of eigenvalues happens at the corresponding higher-order crossing points in the dispersion relation. This is the very well known effect which leads to formation of conducting bands in metals. 

In case of the double-well potential \eqref{eq:double-well_potential} this yields interesting results when $k=2 p_0$, see Fig.\,\ref{fig:double-well2}. Indeed, we already know that $D_{p_0} = D_{-p_0}$, hence when one adjusts the wave-vector of the background potential to be twice the wave-vector of the spontaneously modulated ground-state, the energy of this ground-state \eqref{equ:ground_state_energy} is lowered due to the eigenvalue repulsion: $\varepsilon = 2 m - A/(4m)$. The wave-vectors of this kind we call first order commensurate. We see that the commensurate background lattice provides additional stability to the modulated state so we expect the increase of the critical temperature of the spontaneously formed holographic striped state at the commensurate points.

One can also consider the potential with the wave-vector which is close but not exactly at the commensurate point: $k = 2 p_0- 2\delta p$. The mode crossing point $D_{\tilde{p}-k} = D_{\tilde{p}}$ is then located a little away from the minima of the dispersion relation at $\tilde{p} = p_0 - \delta p$.  The dispersion can in this case be approximated by \eqref{equ:mass}. The energy of the resulting state is estimated accordingly:
\begin{equation}
\varepsilon(p_0 - \delta p) =  2 m + \frac{\delta p^2 }{2 m} - \frac{A}{4 m}.
\end{equation}
If $\delta p^2 < A/2$ this energy is lower then $\varepsilon_0$, the energy of the initial ground state. The state with the wave-vector commensurate with the lattice becomes preferred over the state with spontaneous $p_0$. This effect is called {\it commensurate lock-in} of periodic structures. We see, that for given amplitude $A$ there exists a region $p \in (p_0 - \sqrt{A/2}, p_0 + \sqrt{A/2})$ where the background potential forces the spontaneous structure to assume the wave-vector which is commensurate with the lattice. 

Finally, we would like to point out one more phenomenon, which is observed when the wavelength of the potential becomes large, see Fig.\,\ref{fig:double-well3}. In this case $k \rar 0$ and the crossing points are located in the vicinity of the same minimum of the double-well potential: $D_{p_0 - k/2} = D_{p_0 + k/2}$. The associated splitting leads to 
\begin{equation}
\label{eq:zero_k}
\varepsilon(p_0 - k) = 2m + \frac{k^2 }{2 m} - \frac{A}{4 m},
\end{equation}
and similarly to the previous considerations, the critical temperature is enhanced in the region were $k < \sqrt{A/2}$. One can understand this feature in the following way.  At very small $k \ll p_0$ the scalar field sitting on top of the cosine wave perceives it as a constant background. One can use the adiabatic approximation and consider the effect of the potential as a slow variation of the mass: $m(x) = \sqrt{m^2 - A(x)/4}$. Thus, the ground-state energy will slowly change in space. This in turn can be represented as the effective potential in position space which forces the scalar to roll to the region were the ground state energy is the lowest, i.e. $\cos(k x) = -1$ and the mass is modified as $4 m^2 \rar 4 m^2 - A$ \eqref{eq:mode_coupling}. At $k=0$ this gives the same result for $\varepsilon$ as \eqref{eq:zero_k}. For small $k$ the positive contribution to $\varepsilon$ proportional to $\frac{k^2}{2 m}$ comes from the kinetic energy of the particle sitting in the potential well of 
size $\sim 1/k$. 

At the end of the day, we observe 3 types of commensurate phenomena in our toy model which we summarize in the following propositions:

\begin{theorem}
\label{pr:stability}
Additional stability of the spontaneously modulated phase is provided by the commensurate background potential when $k = 2 p_0$.
\end{theorem}

\begin{theorem}
\label{pr:locking}
Commensurate lock-in between background and spontaneous wave-vectors in the interval of $k$ around the commensurate point. The size of the interval is proportional to the amplitude.
\end{theorem}

\begin{theorem}
\label{pr:long_wave}
Decrease of the ground state energy of the spontaneous mode on top of a long-wavelength potential.
\end{theorem}

We should also emphasize the key features of the model which lead to the above mentioned effects. Firstly, the interaction of the spontaneous structure with the potential is quadratic $\delta \mathit{L} = \phi(x)^2 V(x)$. This is the leading power of $\phi(x)$ in a perturbative series
of a given background. 
%
The effects which we discuss strongly rely on the mixing between linear modes and as such they require the existence of degenerate minima of the perturbation dispersion relation for the decease of ground state energy to happen. For given minima occurring at points $p_1$ and $p_2$, the commensurate effect would be observed at $k = p_1 - p_2$. Because the model under consideration possess a P-symmetry, the minima occur at $p_1 = p_0, p_2=-p_0$, and we get the result $k=2 p_0$ for the leading commensurate point. At finite values of the amplitude, higher order commensurate points $n k = 2 p_0$ should be observed as well.


\section{The model}
\label{sec:the_model}

As mentioned in the Introduction, the background solution we consider is the non-homogeneous ionic lattice of \cite{Horowitz:2012gs} which results from giving spatial modulation to the chemical potential in the Einstein-Maxwell theory in 4-dimensional $AdS$. In order to describe the spontaneous translational symmetry breaking we will consider the model of \cite{Donos:2011bh} which couples the Einstein-Maxwell system to a pseudo scalar via a Chern-Simons (CS)-like term in the action. Following the conventions of \cite{Withers:2013kva, Withers:2013loa}, we write the action as 
\begin{equation}\label{S_0}
	S = \int d^4 x \sqrt{- g} \left( R - \frac{1}{2} (\partial \psi)^2 - \frac{\tau(\psi)}{4} F^2 - V(\psi) \right)
	 - \frac{1}{2} \int {\vartheta}(\psi) F \wedge F.
\end{equation}
Here $F$ is the field strength associated to the Maxwell field ${\cal A}$ and $F = d{\cal A}$. The various couplings in  
the Lagrangian can be chosen in a number of different ways, but we will study the specific case
\begin{equation}
\label{eq:potentials}
	V = - 6 \cosh (\psi /\sqrt{3}) , \qquad \tau = {\rm sech} (\sqrt{3} \psi), \qquad \vartheta = \frac{c_1}{6 \sqrt{2}} \tanh(\sqrt{3} \psi), \qquad c_1 =9.9.
\end{equation}
This choice is motivated by the fact that one can obtain similar couplings from dimensional reduction of supergravity, 
and has been previously considered in \cite{Donos:2011bh, Withers:2013kva, Withers:2013loa, Donos:2013wia}, with which 
our results agree in the pertinent limits. 
Note that in these conventions the cosmological constant is $\Lambda = - 3$ and the mass of the scalar is $m^2 = -2$.

In case of constant chemical potential $\mu = \bar \mu$, the equations of motion admit the translational invariant RN solution which can be written as
\begin{equation}\label{RN soln}
	ds^2 = \frac{1}{z^2} \left( - f(z) dt^2 + \frac{dz^2}{f(z)} + dx^2 + dy^2 \right) , \quad {\cal A} = \bar \mu (1 - z) dt , \qquad \psi = 0,
\end{equation}
\noindent where
\begin{equation}\label{f RN}
	f = (1-z)\left( 1 + z + z^2 - \bar \mu^2 z^3 /4 \right).
\end{equation}
In units of inverse horizon radius the temperature of this solution reads
\begin{equation}\label{eq:T}
 	\boldsymbol{\mathit{T}} = \frac{12 - \bar \mu^2}{16 \pi}
\end{equation} 
and the entropy density is $s = 4 \pi$. Note that in our coordinates the conformal boundary is located at $z=0$ while the 
black hole horizon is at $z=1$.

In order to break the translation symmetry explicitly, we choose the chemical potential to have a periodic dependence in $x$, $	A_t(z=0,x) = \mu(x)$ with 
\begin{equation}\label{eq:mu x}
	\mu(x) = \mu_0 (1 + A \cos(\boldsymbol{\mathit{k}} x) ).
\end{equation}
As stated in \cite{Withers:2013kva, Withers:2013loa, Donos:2013wia}, in order to consider solutions with spatial modulation in the $x$-direction, it suffices to use the ansatz 
\begin{align}\label{ds2 anstaz}
	ds^2 &= \frac{1}{z^2}\left( - Q_{tt} f(z) dt^2 + Q_{zz} \frac{dz^2}{f(z)} + Q_{xx} (dx + Q_{zx} dz)^2 + Q_{yy} ( dy + Q_{ty} dt )^2  \right), \\
\notag
	{\cal A} &= A_t dt + A_y dy 
\end{align}
Here, $Q_{tt}$, $Q_{zz}$, $Q_{xx}$, $Q_{yy}$, $Q_{ty}$, $Q_{zx}$, $A_t$, $A_y$ and $\psi$ are functions of the holographic coordinate $z$
and the boundary coordinate $x$. In the normal state above the critical temperature, the pseudo-scalar $\psi$ vanishes along with $Q_{ty}$ and $A_y$, in which case the ansatz reduces to the one used in \cite{Horowitz:2012gs} for pure ionic lattices.

Without loss of generality, we set $\mu_0 = \bar \mu$ in \eqref{eq:mu x}, 
and use $\bar{\mu}$ as a unit to measure all the dimensionful parameters in the model, denoted up to now with bold script. In what follows we substitute these parameters by the dimensionless ratios
\begin{equation}
\boldsymbol{\mathit{T}} \rar  T \bar{\mu}, \qquad \boldsymbol{\mathit{k}} \rar k \bar{\mu}, \qquad \boldsymbol{\mathit{p}} \rar p \bar{\mu}.
\end{equation}
All our non-linear background solutions will be constructed 
using the DeTurk trick as explained in \cite{Headrick:2009pv, Adam:2011dn, Wiseman:2011by}. 
The stationary solutions satisfy regularity at the horizon, which can be implemented by setting $Q_{tt}(1,x) = Q_{zz}(1,x)$ in addition to 
mixed boundary conditions for the remaining free functions, see e.g. \cite{Horowitz:2012ky}. The condition $Q_{tt}(1,x) = Q_{zz}(1,x)$ ensures 
that the surface gravity is constant so that the temperature of the spatially modulated black branes is given by \eqref{eq:T}.
At the UV, we require the metric to be that of AdS.

\begin{figure}[ht]
\centering
\includegraphics[width=0.6 \linewidth]{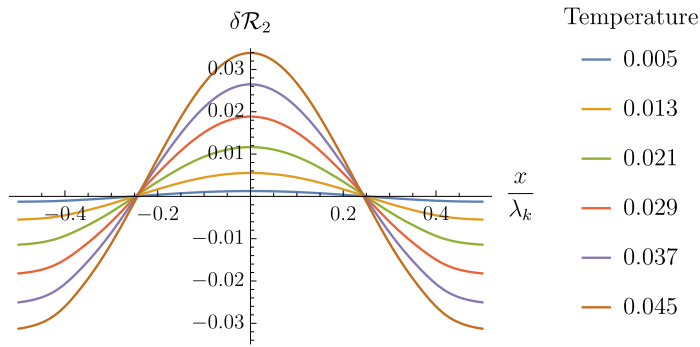}
\caption{\label{fig:Hcurve} Profiles of the relative modulation of curvature on the horizon for the ionic lattice backgrounds at different temperatures.}
\end{figure}

As mentioned above, the formation of the stripes occurs in the near horizon, so it is useful to understand this region 
of the background geometry. It has been shown in \cite{Donos:2014yya} that the ionic lattice approaches the translational invariant $AdS_2 \times R^2$ 
solution at extremality $T \rar 0$. Increasing the temperature, the horizon acquires modulation and this will affect our marginal modes. 
In order to capture this effect, we use as a measure the scalar quantity\footnote{Note that the Ricci scalar is constant because the electromagnetic stress tensor is traceless in four dimensions.}
\begin{equation}
	{\cal R}_2 = R_{\mu \nu} R^{\mu \nu} |_{{\rm horizon}}
\end{equation}
\noindent evaluated at the horizon. To check for horizon modulation, we consider the normalized difference of 
${\cal R}_2$ with respect to its mean value,
\begin{equation}
  \delta {\cal R}_2= 	\frac{{\cal R}_2 - \bar {\cal R}_2 }{\bar {\cal R}_2}
\end{equation}  
\noindent where the bar denotes the spatial average
\begin{equation}
	\bar {\cal R}_2 = \lambda_{k}^{-1} \int_{0}^{\lambda_{k}} {\cal R}_2 (x) dx 
\end{equation}
We show in Fig.\,\ref{fig:Hcurve} the values of $\delta \mathcal{R}_2$ for varying temperature at $k=1.8$ and $A=1.4$. We see that for small temperatures
the horizon modulation is small, consistent with the results of \cite{Donos:2014yya}. Increasing the temperature, the modulation becomes 
significant. 

\section{Marginal modes}
\label{sec:marginal_modes} 

The spontaneous formation of the new spatially modulated striped phase is signalled 
by the appearance of marginal modes with zero frequency and finite momentum $p$ in the spectrum of linear fluctuations around the background solution at certain temperature $T$. In the case when translations are not broken explicitly the normal phase is the homogeneous RN black hole and the study of the marginal modes results in a characteristic ``bell''-shaped instability curve in the $(p,T)$-plane \cite{Donos:2011bh}. 
The tip of the bell, which has maximum temperature $T=T_\mathrm{max}$, occurs at the preferred momentum $p_\mathrm{c}$, where ``the most unstable mode'' forms and drives the system to the new spatially modulated state. As it was shown in \cite{Donos:2013wia,Withers:2013loa,Withers:2013kva}, at lower temperature the thermodynamically preferred nonlinear solution has a wavelength which is slightly different from $2\pi/p_\mathrm{c}$, but nonetheless one can say that $p_\mathrm{c}$ roughly sets the scale of translation symmetry breaking. Hence the study of the perturbative instability curves provides valuable information about the system. 

The values of $T_\mathrm{max}$ and $p_c$ for the homogeneous RN background play an important role in our study, as they characterize the internal (i.e. independent of the lattice) dynamics of the spontaneous wave formation. For the particular model \eqref{eq:potentials} they assume the values:
\begin{equation}
\label{eq:RN_values}
\mathrm{RN}: \qquad T_{\mathrm{max}} \equiv T_{0} \approx 0.024, \qquad p_c \equiv p_0 \approx 0.78
\end{equation}

In what follows we will study the marginal modes with finite momentum on top of the holographic ionic lattice background of \cite{Horowitz:2012gs} with lattice wave-length $\lambda_k$. The background breaks translation symmetry explicitly, so one must take additional care when using the momentum space picture because momentum is no longer a conserved quantity as we have seen already in Sec.\,\ref{sec:physics_of_commensurate_lock}. At the level of the equations of motion we see that the $x$-dependent profiles of the background solution act very much like the periodic potential of \eqref{eq:mode_coupling} and one cannot simply perform a Fourier transform. Instead, when dealing with the linearized perturbations one makes use of the Bloch wave function approach \eqref{eq:Bloch_function}. One can show that it is consistent to turn on the following set of linear perturbations which constitutes the unstable mode\footnote{Remarkably, these are the same fields that need to be 
perturbed in order to capture the marginal modes in the RN background \cite{Donos:2011bh}.}
%
\begin{equation}\label{eq:marginal modes}
	\delta \psi = \delta \psi^p(z, x) e^{i p x}, \qquad \delta A_y = \delta {A_y}^p(z, x) i e^{i p x}, 
	\qquad \delta Q_{ty} = \delta {Q_{ty}}^p(z, x) e^{i p x},
\end{equation}
where now the $x$-dependent profiles have the periodicity $\lambda_k$ of the lattice and $p$ is the Bloch ``pseudo-momentum''. The exponentials can be factored out from the linearized equations of motion and only leave factors of $p$ and $p^2$ behind. 
We emphasize that $p$ being a Bloch pseudo-momentum means that it is only defined up to discrete shifts by the lattice wave-vector $p \sim p + n k, \ n\in \mathbb{Z}$. Indeed, the difference between $p$ and $p+ n k$ in \eqref{eq:marginal modes} can be absorbed into the definition of the corresponding profile function. In what follows it will be convenient in some cases to use the unbounded value $p$ (this accounts to the ``extended BZ''), while keeping in mind this ambiguity in its definition. At the same time, in complete analogy with \eqref{eq:matrix_coupling}, restricting the scope to the pseudo-momenta $\tilde{p}$ inside the Brillouin zone $\tilde{p} \in [0,k)$ is enough to observe any spatially modulated instabilities of the form \eqref{eq:marginal modes} (this is what one can call ``periodic BZ'' representation). 

The fluctuation equations are linear PDE's that possess the schematic form 
\begin{equation}\label{eq:mm}
 {\cal O}^p (z, x) \delta \Phi^p ( z, x ) = 0    
 \end{equation} 
\noindent where ${\cal O}^p (z , x) $ is a second order linear differential operator and $\delta \Phi^p ( z, x )$ collectively denotes the three perturbed fields of \eqref{eq:marginal modes}. Both ${\cal O}$ and $\delta \Phi$
are labeled by $p$, depend on the holographic coordinate $z$, and are periodic in the spatial coordinate $x$ with the period $\lambda_k$. 
Marginal modes are nontrivial solutions to \eqref{eq:mm} with boundary conditions set by the 
regularity at the horizon and vanishing of the sources in the UV. The trivial solution $\delta \Phi^p = 0$ is always present, so we actually need to solve the Sturm-Liouville problem looking for the special values of $p$ which render the linear system underdetermined and allow for nontrivial solutions. 
\begin{figure}[!htbp]
\centering
\begin{subfigure}[t]{0.9\linewidth}
\includegraphics[width = .45\linewidth]{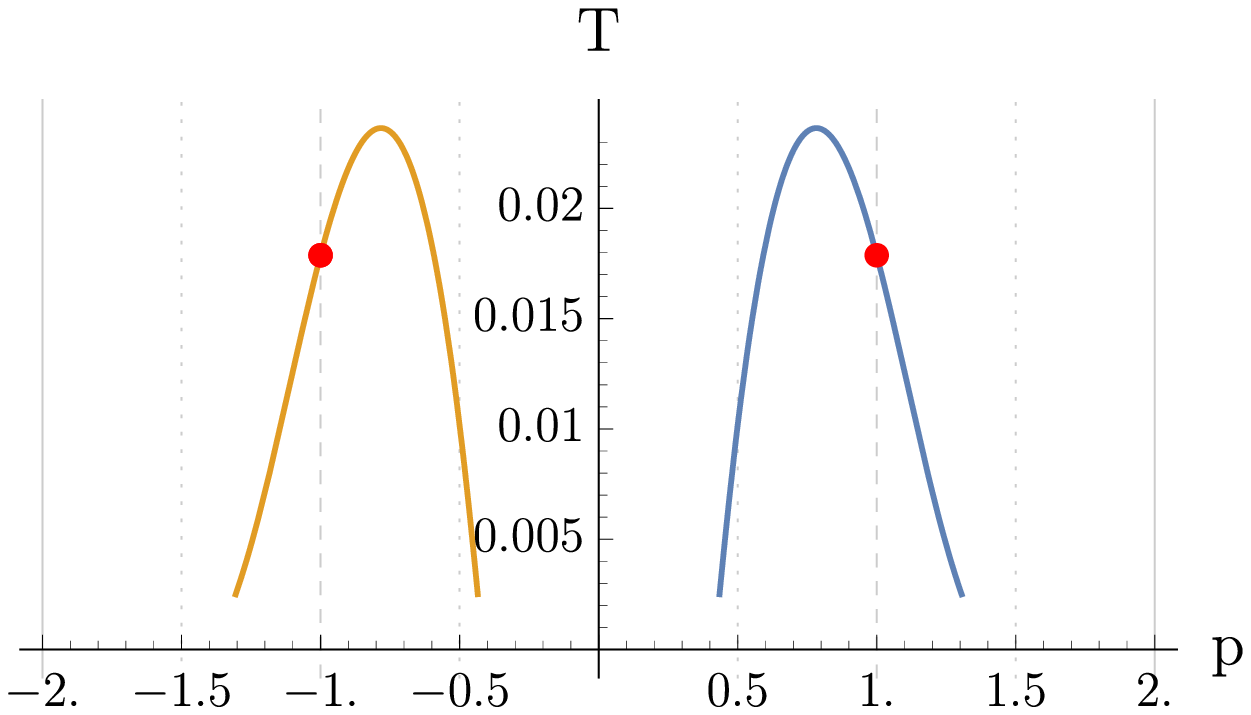}
\qquad
\includegraphics[width = .5\linewidth]{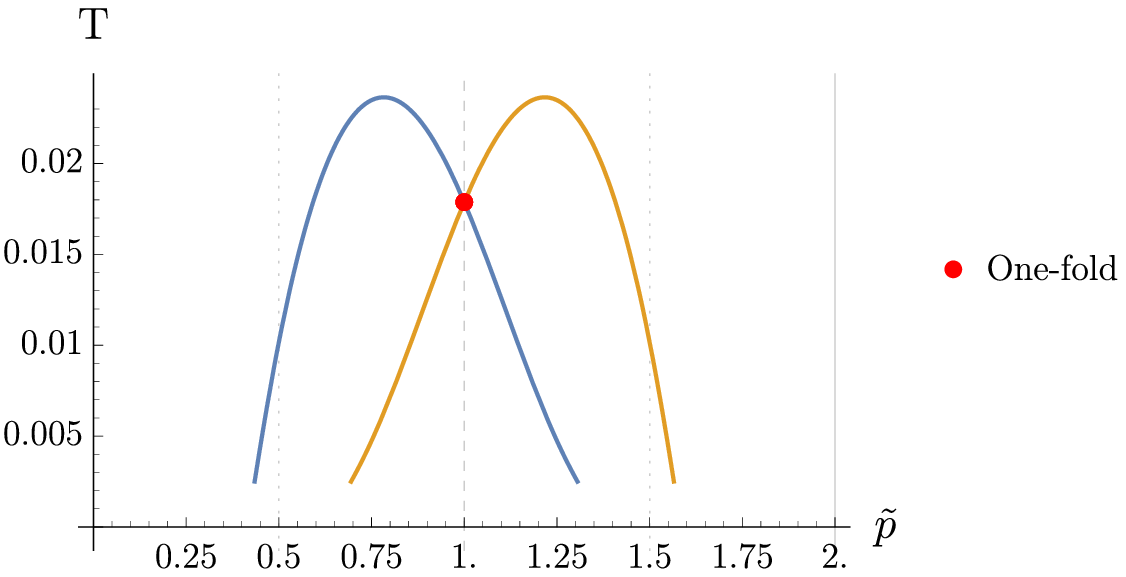}
\caption{Lattice wave-vector is relatively large ($k=2.$), there is only one crossing point in the folded BZ}
\label{fig:FoldingP2}
\end{subfigure}

\begin{subfigure}[t]{0.9\linewidth}
\includegraphics[width = .45\linewidth]{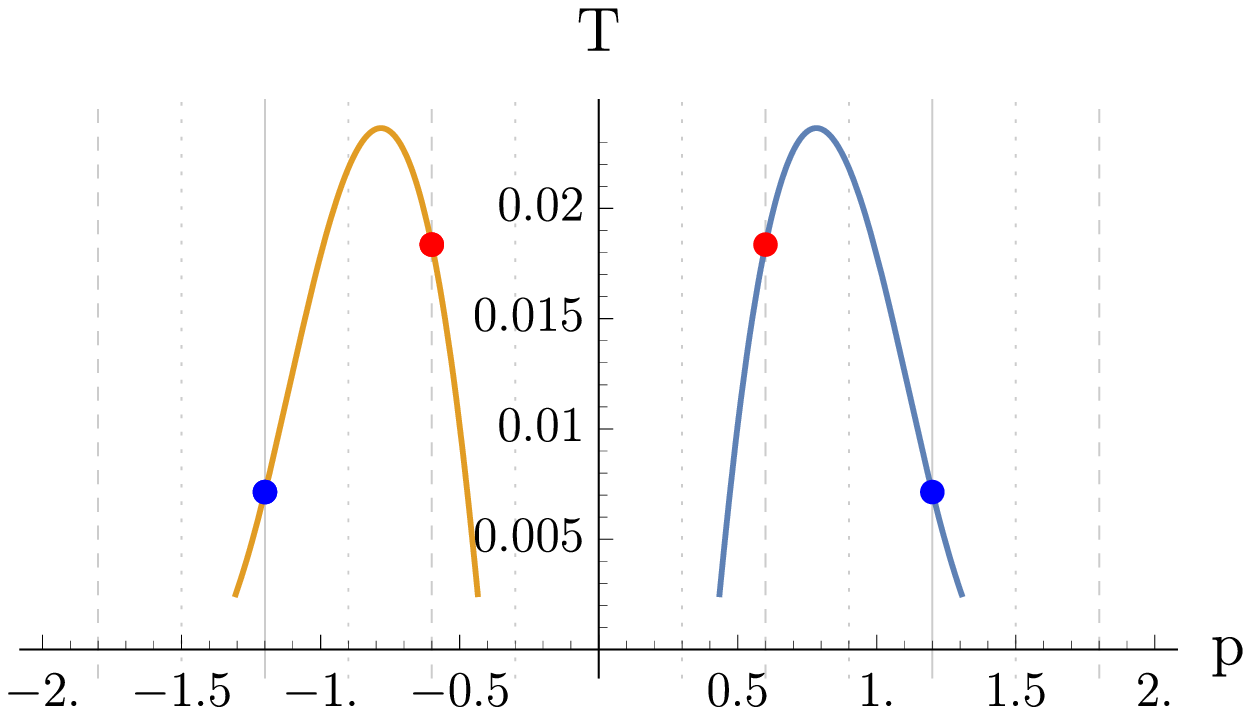}
\qquad
\includegraphics[width = .5\linewidth]{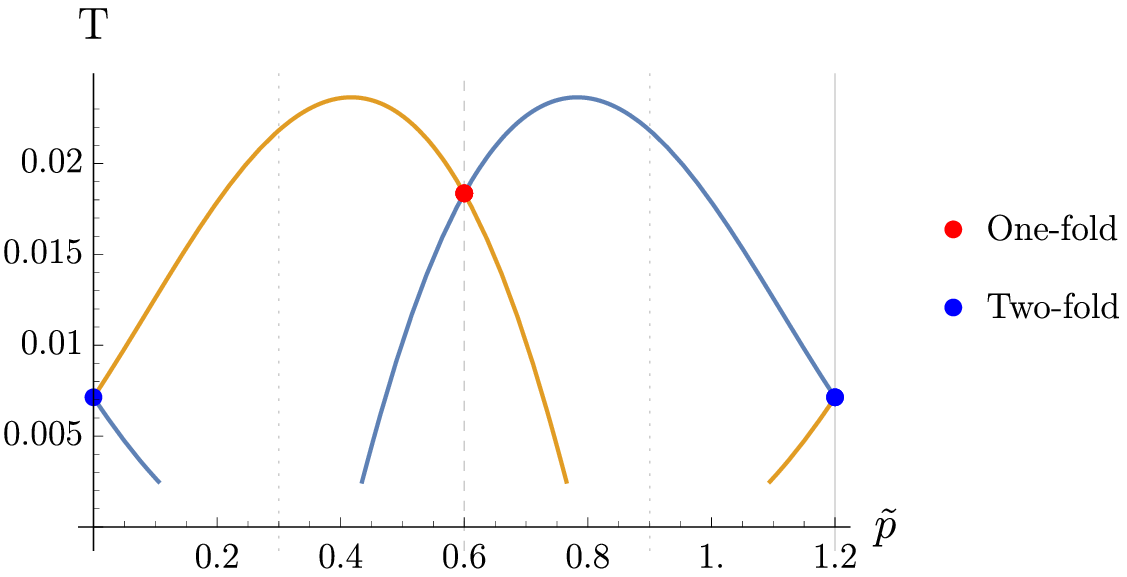}
\caption{When the lattice wave-vector get smaller $k=1.2$, additional crossing points appear.}
\label{fig:FoldingP12}
\end{subfigure}

\begin{subfigure}[t]{0.9\linewidth}
\includegraphics[width = .45\linewidth]{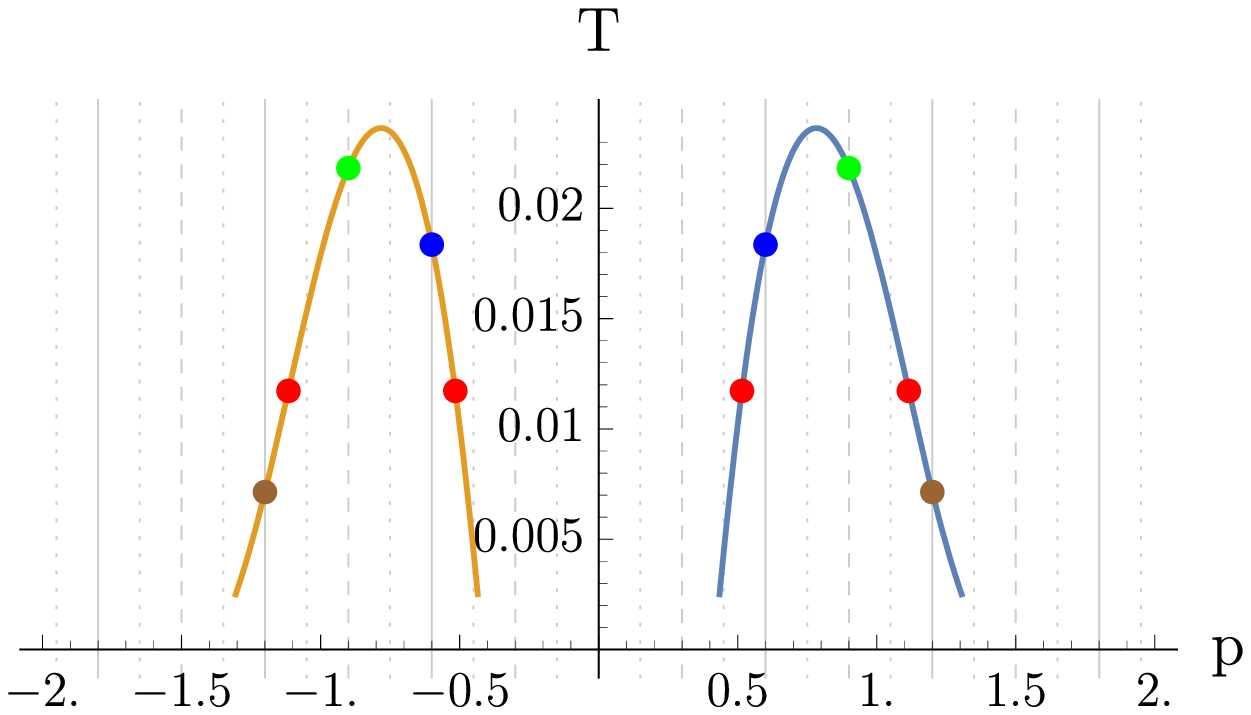}
\qquad
\includegraphics[width = .5\linewidth]{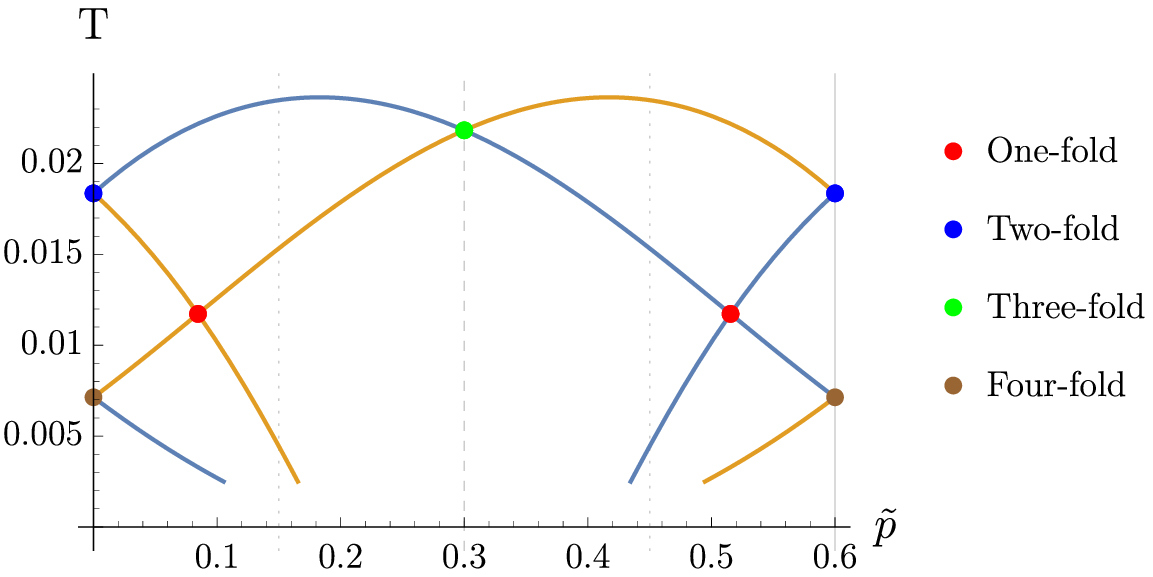}
\caption{When the lattice wave-vector is relatively small ($k=0.6$), higher order foldings are abundant.}
\label{fig:FoldingP06}
\end{subfigure}
\caption{Cartoon of the Brillouin zone folding using the instability curve of RN marginal modes as an example. Left panels show the extended BZ representation, right panels show the Bloch momentum in the periodic BZ at different wavevectors of the ionic lattice. Note that 
n-fold points are separated by distance in momentum space $\Delta p = n k$.}
\label{fig:Folding}
\end{figure}

Note that the ionic lattice background is parity (P) symmetric, 
but this is broken spontaneously by the striped mode due to the CS term in the action \eqref{S_0}. As a consequence of this, any marginal mode has a symmetric counterpart, which one can obtain by applying the parity transformation
\begin{equation}
 \mathrm{P}: \qquad  p \rar -p, \qquad \psi \rar -\psi, \qquad Q_{ty} \rar -Q_{ty}.   
\end{equation}
Given this, all eigenvalues $p$ will appear in pairs in our study and this is why the intuition we developed in Sec.\,\ref{sec:physics_of_commensurate_lock} is relevant in this case. We can get further intuition by inspecting the behavior of the instability curves at vanishing amplitude of the ionic lattice, which we sketch on Fig.\,\ref{fig:Folding}. Indeed, the analogue to the double-well toy model is a couple of the RN  instability curves, related by parity, seen on the left panels of Fig.\,\ref{fig:Folding}\footnote{In the translationally invariant case,
one can obtain these curves by using the shooting method of \cite{Donos:2011bh}. For the specific choice $c_1 = 9.9$, see 
\eqref{eq:potentials}, our results agree with those in \cite{Withers:2013loa}.}. 
Even before we consider finite values of the modulation amplitude $A$ we already see that the $\lambda_k$-periodic lattice profiles will lead to the ``folding" of these curves in the corresponding BZs, as shown on the right panels of Fig.\,\ref{fig:Folding} for different lattice momenta.
We expect the eigenvalue repulsion to happen at the points separated by the integer multiples of $k$. At small $A$ it will be dominated by the one-fold crossings.
As we see on Figs.\,\ref{fig:FoldingP12}, \ref{fig:FoldingP2}, at low lattice wave-vectors ($k \lesssim 1.2$) crossing points of higher order appear in the BZ, and we expect the gaps to open there for finite values of $A$, when higher order of perturbation series are important. 
On the other hand, at larger wave-vectors 
($k \gtrsim 2.5$) the instability curves do not cross within the BZ, thus no interesting commensurate effects should be expected. 

At finite values of the ionic lattice amplitude, the backgrounds satisfy the nonlinear Einstein-Maxwell equations with periodic chemical potential and the perturbation equations \eqref{eq:mm} can be solved by discretizing using a pseudospectral collocation method. %
Upon discretization, ${\cal O}^p$ in \eqref{eq:mm} is a matrix so the condition for the existence of non-trivial solutions is that $p$ is such that 
\begin{equation}\label{eq: eq for mm}
	{\rm det} \,  {\cal O}^p  = 0,
\end{equation}
which is an algebraic matrix equation. 
Generically, this equation will only have complex $p$ solutions, which are inadmissible for us, but below a critical temperature real eigenvalues appear in a similar manner to what happens in the RN case \cite{Donos:2011bh, Withers:2013kva, Withers:2013loa, Donos:2013wia}. 
%
More details of the numerical search for the eigenvalues of \eqref{eq: eq for mm} can be found in the Appendix. 
\begin{figure}[!htbp]
\centering{}

\begin{subfigure}{0.9\linewidth}
\includegraphics[width = 1.\linewidth]{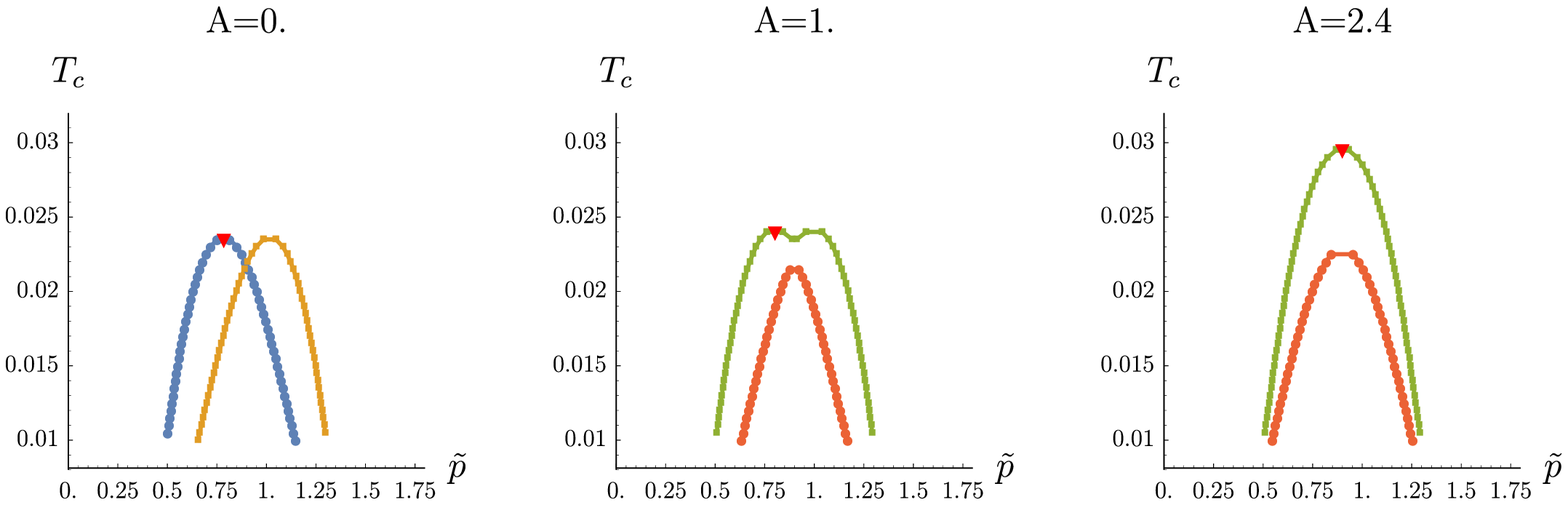}
\caption{\label{fig:bell18} Larger $k = 1.8$, only one crossing point is present.}
\end{subfigure}
\begin{subfigure}{0.9\linewidth}
\includegraphics[width = 1.\linewidth]{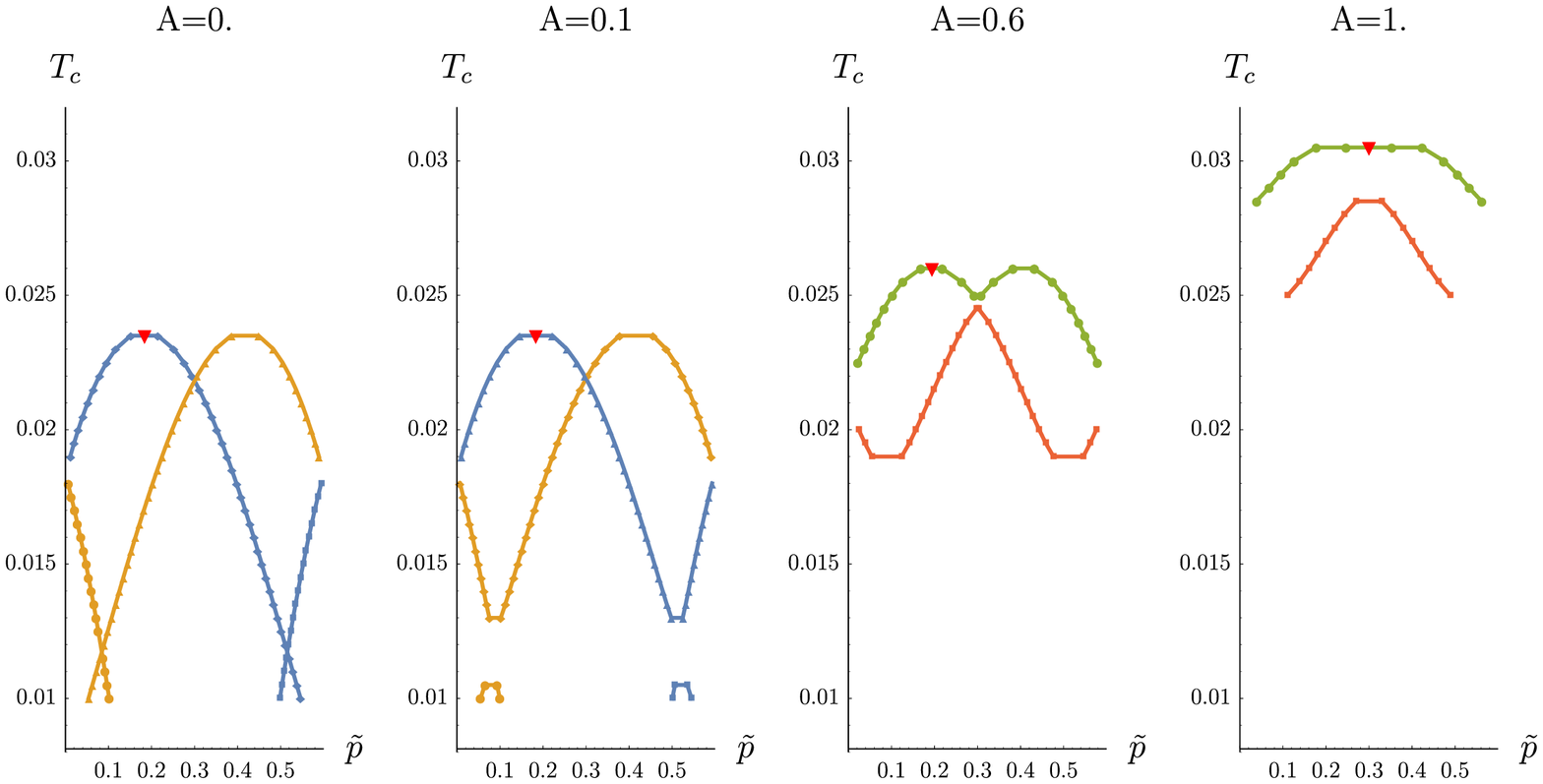}
\caption{\label{fig:bell06} Lower $k = 0.6$, a few gaps are consequently opening in various crossing points.}
\end{subfigure}
\caption{\label{fig:bells} Modified instability curves of the striped marginal mode on top of the ionic lattice background for different lattice wave-vector and amplitude. At large amplitude the commensurate value $\tilde{p}_c=k/2$ has the largest critical temperature.}
\end{figure}

By tracing the position of the real eigenvalues of \eqref{eq: eq for mm} for the series of backgrounds with given period $\lambda_k$, amplitude $A$
and temperature $T$ we obtain the shape of the modified instability curves. As expected, we observe that the coalescing points start interacting and repel each other with a distance that grows with the amplitude of the lattice as it is shown in Fig.\,\ref{fig:bells}, which represents a sample of our data for $k=1.8$ and $k=0.6$. This separation in turn increases the critical temperature of the marginal modes and causes the preferred momentum (shown as a red triangle) to deviate from the RN value, up to a point in which the dominant mode occurs exactly at the commensurate point. Note that the zero amplitude plots compared with Fig.\,\ref{fig:Folding} serve as a good check for consistency of our results.

One important subtlety arises when analyzing the instability curves of Fig.\,\ref{fig:bells}. The plots show the periodic BZ, characterized by the Bloch momentum $\tilde{p}$. We see that in both cases the commensurate point is achieved at $\tilde{p}_c = k/2$, but some more care should be taken when determining its order. Indeed for $k=0.6$, Fig.\,\ref{fig:bell06}, $\tilde{p}_c=0.3$, which is equivalent in the expanded BZ to $p_c=0.3$, or $p_c =0.3 + k=0.9$, etc. These values correspond in turn to the commensurate points $\frac{1}{2} k$ (first order) or $\frac{3}{2} k$ (third order). We can distinguish between them by addressing the BZ folding argument, depicted on Fig.\,\ref{fig:FoldingP06}. It is clearly seen that the highest temperature commensurate point at $\tilde{p}=0.3$ is actually of third order (the two crossing branches are separated by $3k$) and henceforth corresponds to $p_c = \frac{3}{2} k$ in the expanded BZ representation. Similar analysis for $k=1.8$,  Fig.\,\ref{fig:bell18}, shows that in this case the first order commensurate point $p_c =\frac{1}{2} k$ is the leading one. Another way to understand this effect is to recall that it usually takes more ``effort'' for the lattice to drive the wave-vector of the stripe further out of its dynamically preferred value $p_c=p_0$. Hence it is the commensurate point, lying closest to $p_0$, which becomes the leading one.

\begin{figure}[ht]
\centering
\includegraphics[width = .8\linewidth]{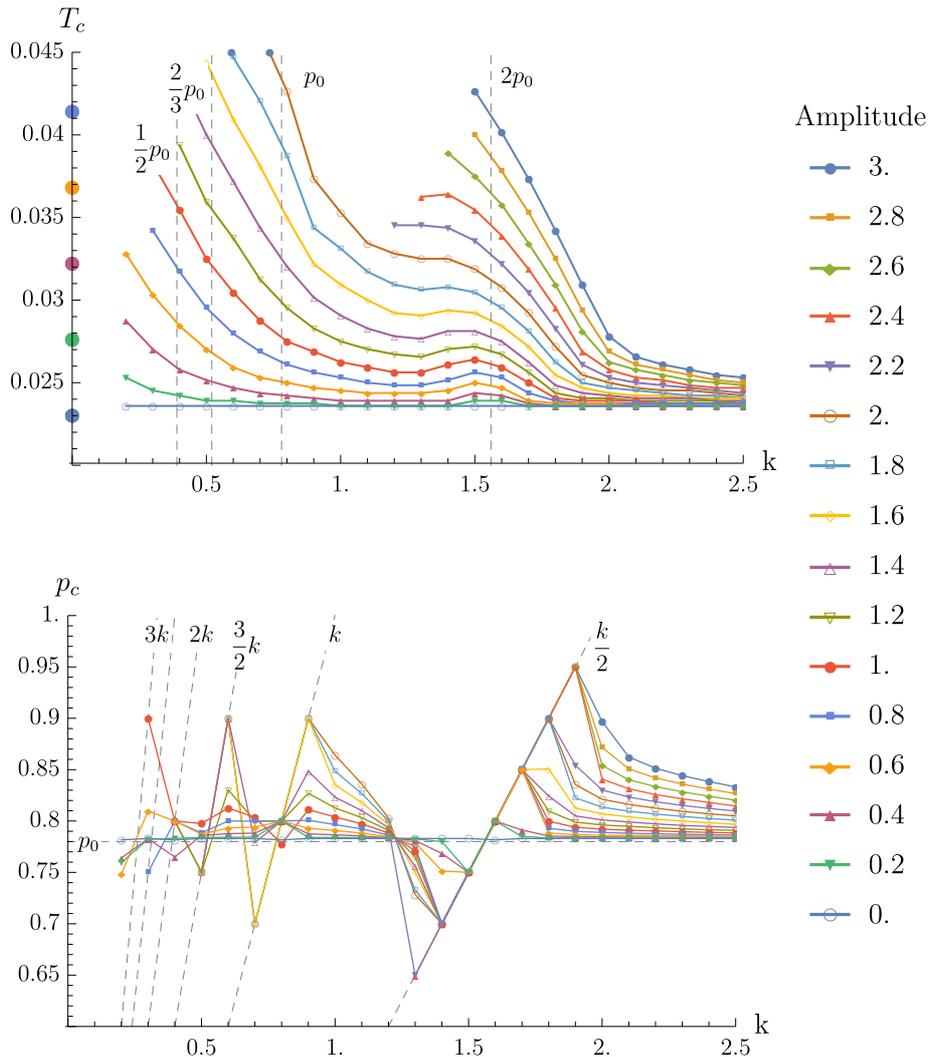}
\caption{Critical temperature and preferred wave-vector for the leading striped instability for the backgrounds with different lattice wave-vector and amplitude. Our data collection is restricted by the precision issues discussed in the Appendix.}
\label{fig:TPplot}
\end{figure}

Having obtained the instability curves for a range of backgrounds with $k \in \{0.2, 0.3, \dots ,2.5\}$ and $A \in \{0.,0.2,\dots, 3.\}$, 
we determine the position of the tip of each curve, which shows the critical temperature $T_\mathrm{max}$ and the wave-vector $p_c$ of the leading unstable stripe mode. Note that, as discussed above, we do not necessarily chose the value of $p_c$ lying inside the first BZ. 
The dependence of these values on $k$ at various $A$ is shown on Fig.\,\ref{fig:TPplot}. This plot is the main result of our study and we discuss it in the next Section.

One comment is in order before we proceed with the discussion in Sec \ref{sec:discussion}. Since the formation of the stripes occurs in the near horizon, one would in principle expect that the marginal modes of ionic lattices, which are known to be  
irrelevant in the IR, will be close to those of RN. This is the case for homogeneous lattices \cite{Andrade:2015iyf}, but as we have seen, this picture drastically changes when we consider non-homogeneous lattices. 
To understand this, recall that the horizon becomes modulated at any non-zero temperature, see Fig.\,\ref{fig:Hcurve}. Since the main feature responsible
for the structure of our marginal modes is simply the periodicity of the lattice (no matter how small the amplitude is) it is reasonable 
to expect that the geometry will affect the marginal modes, specially of the most unstable ones.

\section{Discussion and Summary}
\label{sec:discussion}
 
Let us now discuss the results obtained in Sec.\,\ref{sec:marginal_modes}.
In the plot for critical temperature Fig.\,\ref{fig:TPplot}~(top), we see clear evidence of the commensurate phenomena listed in Sec.\,\ref{sec:physics_of_commensurate_lock}. Indeed, for small amplitudes $A\lesssim 1.4$ the rise of the critical temperature is seen near $k=2p_0$ 
in accordance to Proposition\,\ref{pr:stability}.
Moreover, the critical temperature also rises towards smaller momenta $k \lesssim \sqrt{A} p_0$ and approaches the asymptotic value $T_\mathrm{max}(k=0) = T_0 (1+A)$ (shown by the large dots on the vertical axis), which would be the critical temperature of the unstable mode in a homogeneous background with modified chemical potential $\mu = \bar{\mu}(1+A)$, in agreement with Proposition\,\ref{pr:long_wave}.

\begin{figure}[ht]
\center
\includegraphics[width=0.5 \linewidth]{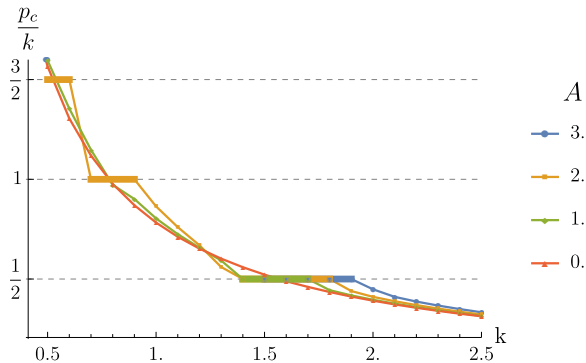}
\caption{\label{fig:Ladder} Same data as Fig.\,\ref{fig:TPplot} rescaled to show the relation between $p_c$ and $k$. The characteristic ``ladder'' shape and the lock-in plateaus (bold lines) are seen.}
\end{figure}

The most striking result is seen on the plot for preferred momentum, see Fig.\,\ref{fig:TPplot}~(bottom).
There we observe, for the first time in holography, that the spontaneous striped wave is forced by the background to assume the lattice wavenumber for strong enough amplitude, as anticipated in Proposition\,\ref{pr:locking} of Sec.\,\ref{sec:physics_of_commensurate_lock}. 
Furthermore, this commensurate lock-in between two periodic structures persists longer for stronger amplitudes, forcing larger deviations of the stripe wave-vector from the preferred one in the homogeneous background. This effect is most pronounced at the first order commensurate point, where the stripe wave-vector locks to $p_c = \frac{k}{2}$, observed in our data in the range $k \in (1.2,2.)$.
The higher order lock-in is seen in our data as well. As it was discussed in Sec.\,\ref{sec:marginal_modes} for the example with $k=0.6$, at smaller lattice wave-vectors the preferred momenta of the stripe align themselves consequently with $p_c = k$ when $k <1.2$, with $p_c = \frac{3}{2}k$ when $k<0.6$ and so on. When plotting the rescaled value $p_c/k$, see Fig.\,\ref{fig:Ladder}, one clearly see the ``ladder'' of lock-in plateaus at different commensurate points. At smaller momenta our sampling in $k$ does not allow us to trace the smooth curves due to rising density of the higher-order commensurate points.
In view of these results, we state that indeed the non-homogeneous lattices in holographic models provide a mechanism to stabilize, pin down and lock the spontaneous modulated structures arising due to horizon instabilities in holographic models.              

\begin{figure}[ht]
\center
\includegraphics[width=0.6 \linewidth]{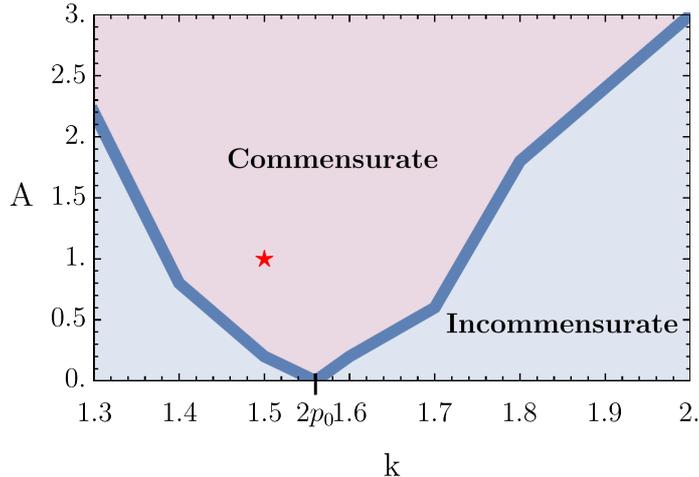}
\caption{\label{fig:PhaseDiag} Phase diagram showing the minimal amplitude of the lattice, which leads to the commensurate lock-in between the spontaneous stripe and the lattice. The range of $k$ corresponds to the first order lock-in, which is seen most clearly in our data. The red star shows the solution depicted in Figs.\,\ref{fig:densityA} and \ref{fig:compare} below.}
\end{figure}

Before the lock-in occurs, we observe another interesting regime when at weaker amplitude the stripe adopts a preferred wave-vector $p_0$ different from $k$. This coexistence of two length scales in the same state leads to very interesting physical phenomena. For instance, the smallest periodic unit cell in this state must simultaneously accommodate integer numbers of two unrelated scales and can thus be very large. In condensed matter, this phenomenon is seen in the context of ``magnetic unit cells" and yields to reconfiguration of the spectrum, folding of Fermi surfaces, etc.

We display in Fig. \ref{fig:PhaseDiag} the phase diagram showing the critical amplitude at which the first order lock-in $p_c = k/2$ occurs as a function of $k$. As expected, it requires stronger lattice to force the stripe wave-length to deviate farther from value preferred by its internal dynamics. Correspondingly, the incommensurate region, where the stripe wave-vector is effectively unrelated to the lattice one can be seen even at considerable amplitudes provided one stays away from commensurate point $k=2 p_0$.

It is worth mentioning that at large $k$ we do not observe commensurability effects in this model, as can be seen in Fig.\,\ref{fig:TPplot}
where both the critical temperature and momentum approach the corresponding translational invariant values. As discussed above, this can be explained noting that 
at large lattice momenta the RN instability curves do not overlap in the BZ so there is no mixing of modes.

Summarizing, we have found that the formation of spontaneous stripes on the holographic 
ionic lattice is sensitive to the commensurability of the wave vectors $k$ and $p$ characterizing the modulation of these structures. 
This phenomenon takes place even though the lattice is 
irrelevant in the IR at zero temperature, and to a large extent can be understood simply from the periodic identification of the Bloch momentum of the striped modes. 
Our analysis is mainly perturbative, consisting of the study of the marginal modes which indicate the instability of the
lattice towards stripe formation. 
We see that the lattice causes the eigenvalue repulsion phenomenon in the instability curves of the striped marginal modes, which overlap in the BZ. This makes the coalescing modes move away from each other and shifts the critical momentum $p_c$ until a point in which it becomes commensurate with the lattice momentum $k$. This lock-in is accompanied by a rise of the critical temperature $T_c$, characterizing the transition to the striped ground state.

At this point it is tempting to discuss the possible physical properties of the nonlinear spatially modulated solutions which would be the end point of the instability triggered by the marginal modes in the commensurate lock-in regime. Conceptually, one can think of the ionic lattice as a UV effect, which introduces explicit translation symmetry breaking, but is suppressed in the IR due to the specifics of the renormalization group flow. Its effect on the IR physics diminishes at lower temperatures, which culminates by its irrelevancy at $T=0$. The spontaneous stripe, on the contrary, is an IR effect. It is sourced by a dynamical instability near the horizon and henceforth has always strong impact on the IR physics. The stripe though does not break the translations explicitly and therefore does not lead to momentum dissipation. Henceforth the commensurate lock-in between the stripe and the lattice has striking consequences. Due to the nonlinear interaction in the bulk between these UV and IR structures, the explicit translational symmetry breaking is conveyed all the way to the IR and benefits from the presence of the stripe near the horizon. In a nut shell we observe the qualitative change from irrelevancy to relevancy of explicit translational symmetry breaking in IR due to the lock-in. One would expect the features of this commensurate state to be similar to that of the IR relevant explicit holographic lattices, which due to the algebraic suppression of DC conductivity at low temperatures are sometimes called ``algebraic insulators''. Hence the commensurate lock-in in our model turns the IR irrelevant metallic state into the IR relevant ``algebraic insulating'' state. This is conceptually similar to the physics of a Mott insulator, described in the Introduction. One has to keep in mind though, that so far there are no examples of the gapped insulating state in holography, which would have exponential suppression of DC conductivity at low temperature, so the analogy with the real world Mott insulator, which is gapped, can only be seen as far as the physics of commensurability is concerned.

\begin{figure}[!tbp]
  \centering
  \begin{minipage}[b]{0.45\textwidth}
  \centering
    \includegraphics[width=1.\textwidth]{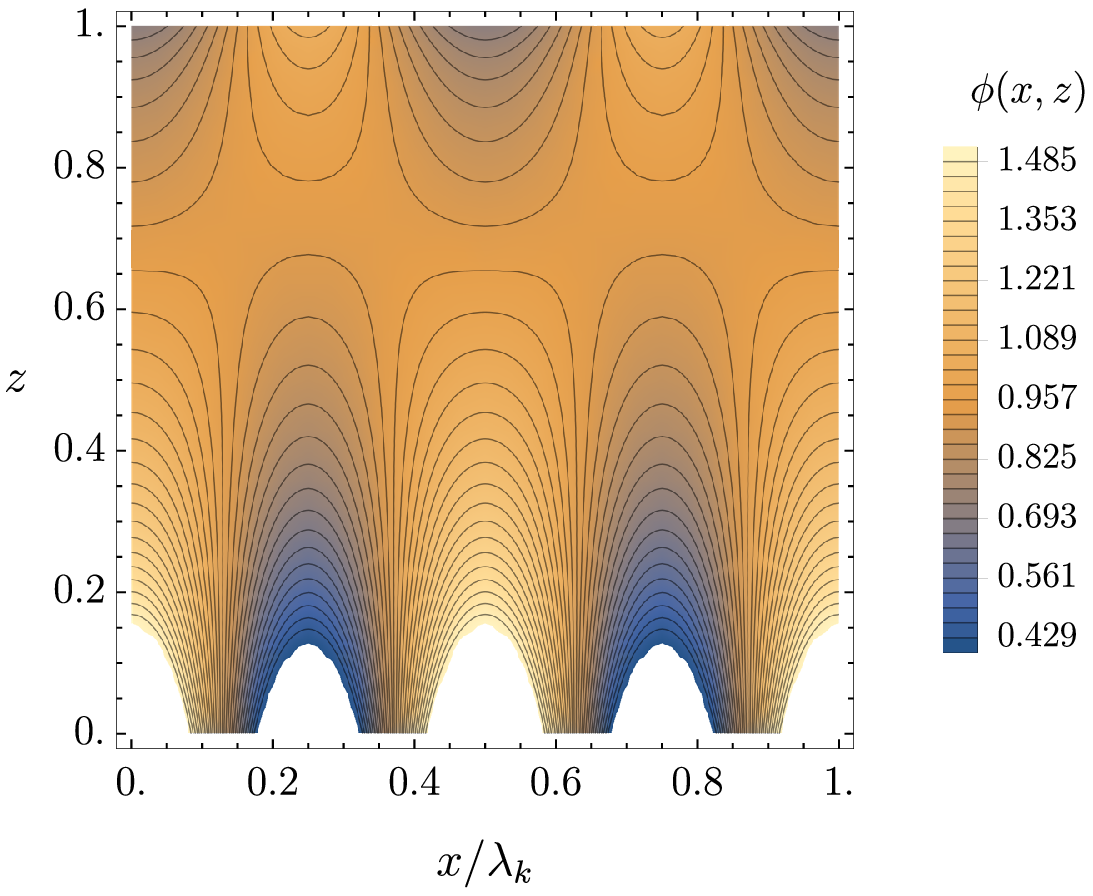}
    \caption{\label{fig:densityA} Nonlinear solution in the commensurate phase for 
    $A=1, \, k=1.5$ and low temperature $T=0.01$. The profile of $\phi(x,z)={A_t(x,z)}/{\bar{\mu} f(z)}$ is shown.} 
  \end{minipage}
\qquad
  \begin{minipage}[b]{0.45\textwidth}
  \centering
    \includegraphics[width=0.9\textwidth]{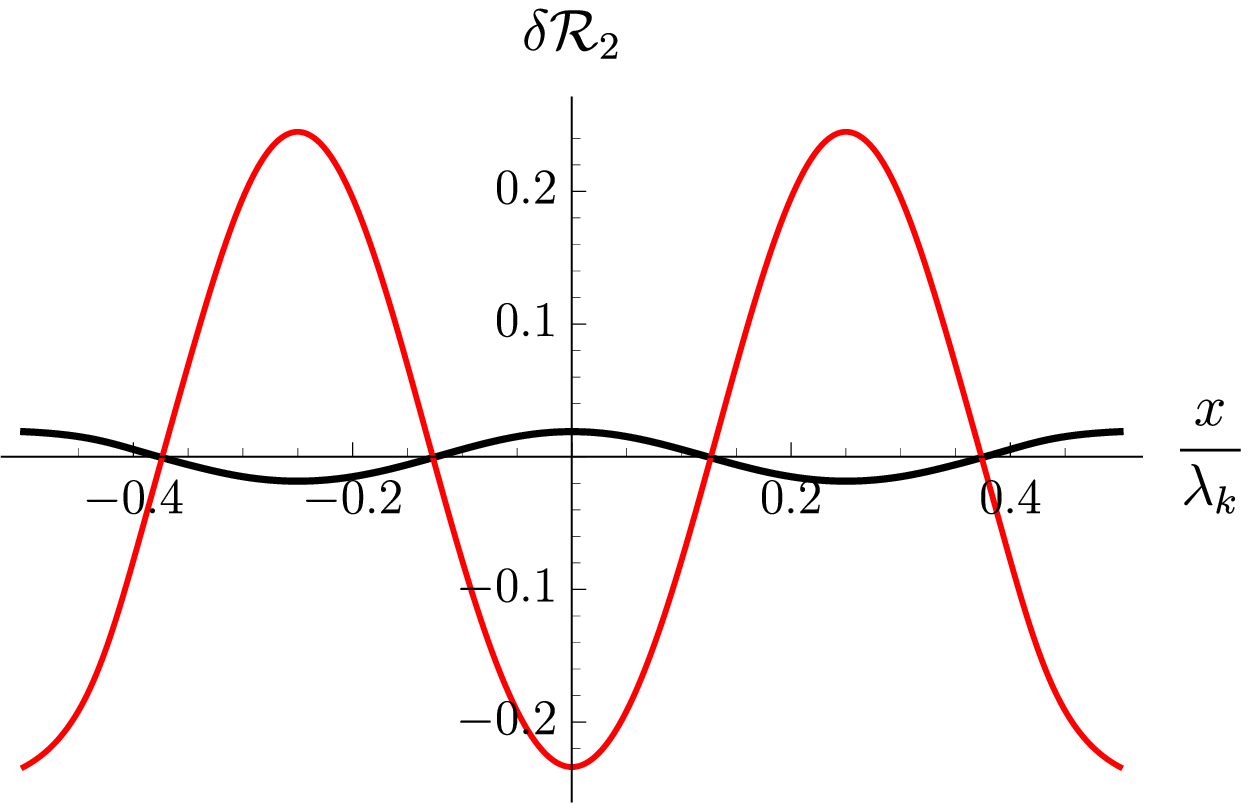}
    \\[1.5cm]
    \caption{\label{fig:compare} Relative curvature modulation for the nonlinear solution with stripe (red curve) and without stripe 
    (black curve).}
  \end{minipage}
\end{figure}

We can get some preliminary signs of the described behavior by constructing the non-linear solution including both the stripe and the lattice in the region where the commensurate lock-in between them is expected to occur, i.e. $k=1.5$, $A=1$, see Fig.\,\ref{fig:PhaseDiag}. We do so by starting with a non-linear solution corresponding to a spontaneous stripe without the lattice
with a given momentum $p = k/2$\footnote{See \cite{Withers:2013kva, Withers:2013loa, Donos:2013wia} for the construction of these non-linear solutions.}, and then adiabatically turning 
on the lattice up to the desired amplitude. 
The profile of a representative solution is seen on Fig.\,\ref{fig:densityA}. We plot the function $\phi(x,z) = \frac{A_t(x,z)}{\bar{\mu} f(z)}$, see \eqref{ds2 anstaz}. It asymptotes to the normalized profile of chemical potential at $z=0$ and to the near-horizon electric field at $z=1$. It appears that the lock-in is mediated by the mutual repulsion of the two periodic structures.
We can clearly see the enhancement of the modulation of the curvature near horizon by comparing the value $\delta \mathcal{R}_2$ for the two solutions with identical ionic lattice sources, but either with and without spontaneous striped structure in the IR, as shown on Fig.\,\ref{fig:compare}. The difference in modulation amplitude by about a factor of 10 provides a clear support to the above argument.

It is interesting to perform a detailed study of the expected holographic commensurately locked state, which would be the endpoint of the instability studied in this work. The salient feature of it would be the commensurately alternating magnetization currents, commensurate charge density wave and relevant translational symmetry breaking in the IR.
While some of the features of the locked-in state can also be described by the simpler models of the explicit IR relevant lattices, the present study opens the possibility to access even more interesting physics which arises when one considers the phase transition to the incommensurate state. The corresponding phenomena would include incommensurate stripes, discommensuration lattices, devil staircases etc. 

It is yet to be checked what state, commensurate or incommensurate, is thermodynamically preferred at given temperature and in a given region of parameter space. In order to study this issue one can not rely on the perturbative analysis of our Fig.\,\ref{fig:PhaseDiag}, Instead, the construction of the fully nonlinear solutions and the study of the free energy of the competing phases at fixed temperature is needed. 
We will come back to these interesting questions elsewhere.

\acknowledgments
We appreciate the contribution of Nikolaos Kaplis at the early stages of this project. We are grateful to Jan Zaanen and Koenraad Schalm for valuable comments and suggestions. We thank Aristomenis Donos, Blaise Gout\'eraux, Simon Gentle and Christiana Pantelidou for useful discussions. 

The work of T.A. is supported by the European Research Council under the European Union’s Seventh Framework Programme (ERC Grant agreement 307955). 
The work of A.K. is supported in part by the VICI grant of Koenraad Schalm from the Netherlands Organization for Scientific Research (NWO), by the Netherlands Organization for Scientific Research/Ministry of Science and Education (NWO/OCW) and by the Foundation for Research into Fundamental Matter (FOM). A.K. is partially supported by RFBR grant 15-02-02092a. 

We are grateful to the organisers of the conference “Numerical Relativity and Holography” held in Santiago de Compostela in July 2016 for their hospitality during the completion of this project.

Most of the numerical computations of this work were carried out in the infrastructure of Instituut Lorentz Workstations.

\appendix

\section{Precision issues in numerical search of marginal modes}
\label{app:precision}
As it has been discussed in Sec.\,\ref{sec:marginal_modes}, upon discretization of the perturbation equations of motion on top of the given ionic lattice background, the search marginal modes reduces to finding the eigenvalues of the matrix ${\cal O}(p)$ in \eqref{eq: eq for mm}. This is a generalized eigenvalue problem, but due to the fact the $p$ appears only up to quadratic order, it can be reduced to the standard linear one. We use \texttt{Eigenvalues} function of Wolfram Mathematica \cite{Mathematica10} for this task. The number of the eigenvalues is, of course, set by to the size of the corresponding matrix, which depends on the size of the grid used for discretization. More precisely, if we use the grid of $N_x \times N_z$ points, we should get exactly $N_x N_z$ eigenvalues, not necessarily different. On the other hand, as it was discussed earlier, if certain $p^*$ is an eigenvalue, then any $p^*+n k$ is eigenvalue as well for any integer $n$. So in the continuous problem one would expect the eigenvalues to arise in infinite sets. We clarify this controversy in this Appendix.

\begin{figure}[!ht]
\centering
\begin{subfigure}[t]{0.45\linewidth}
\includegraphics[width = \linewidth]{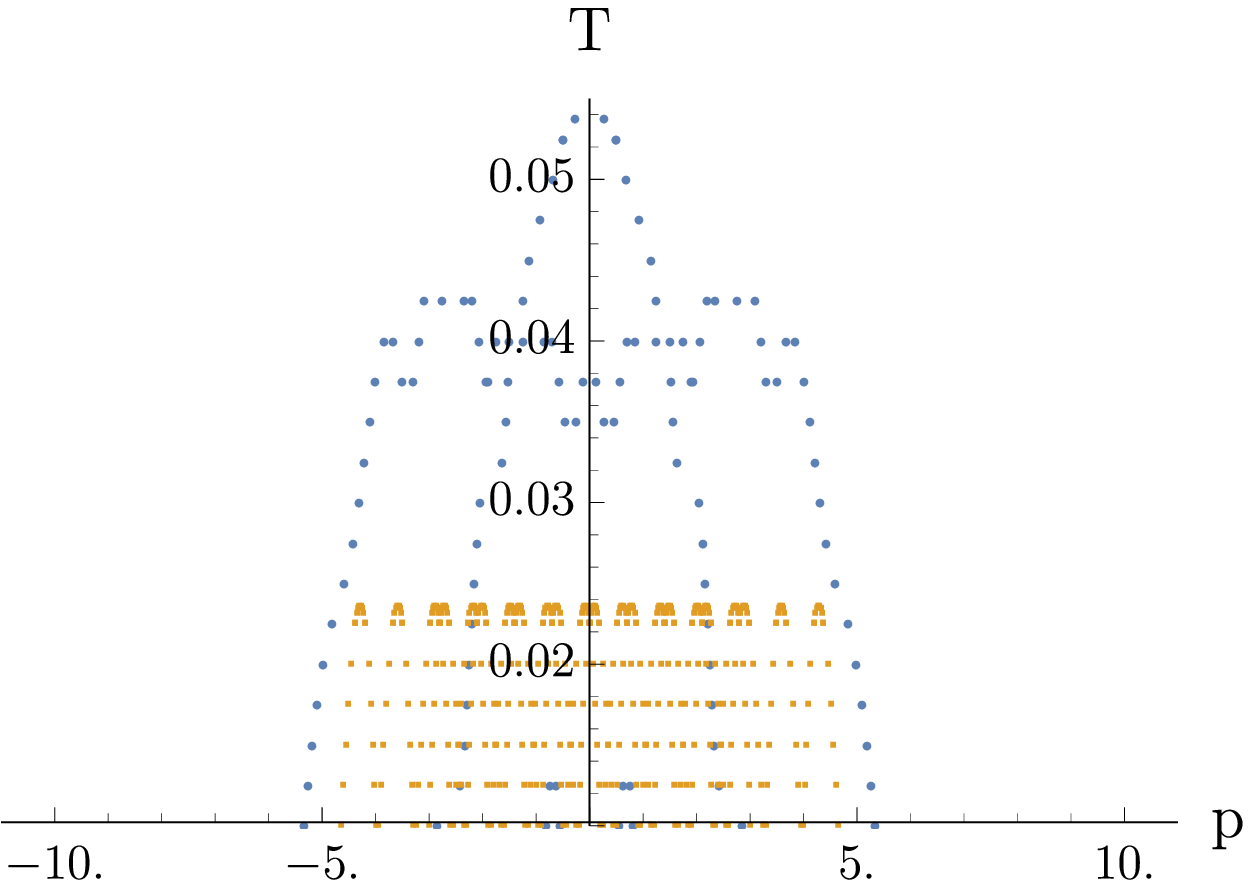}
\caption{\label{fig:LowPrecBell}
Lower resolution $N_x = 17$}
\end{subfigure}
\qquad
\begin{subfigure}[t]{0.45\linewidth}
\includegraphics[width = \linewidth]{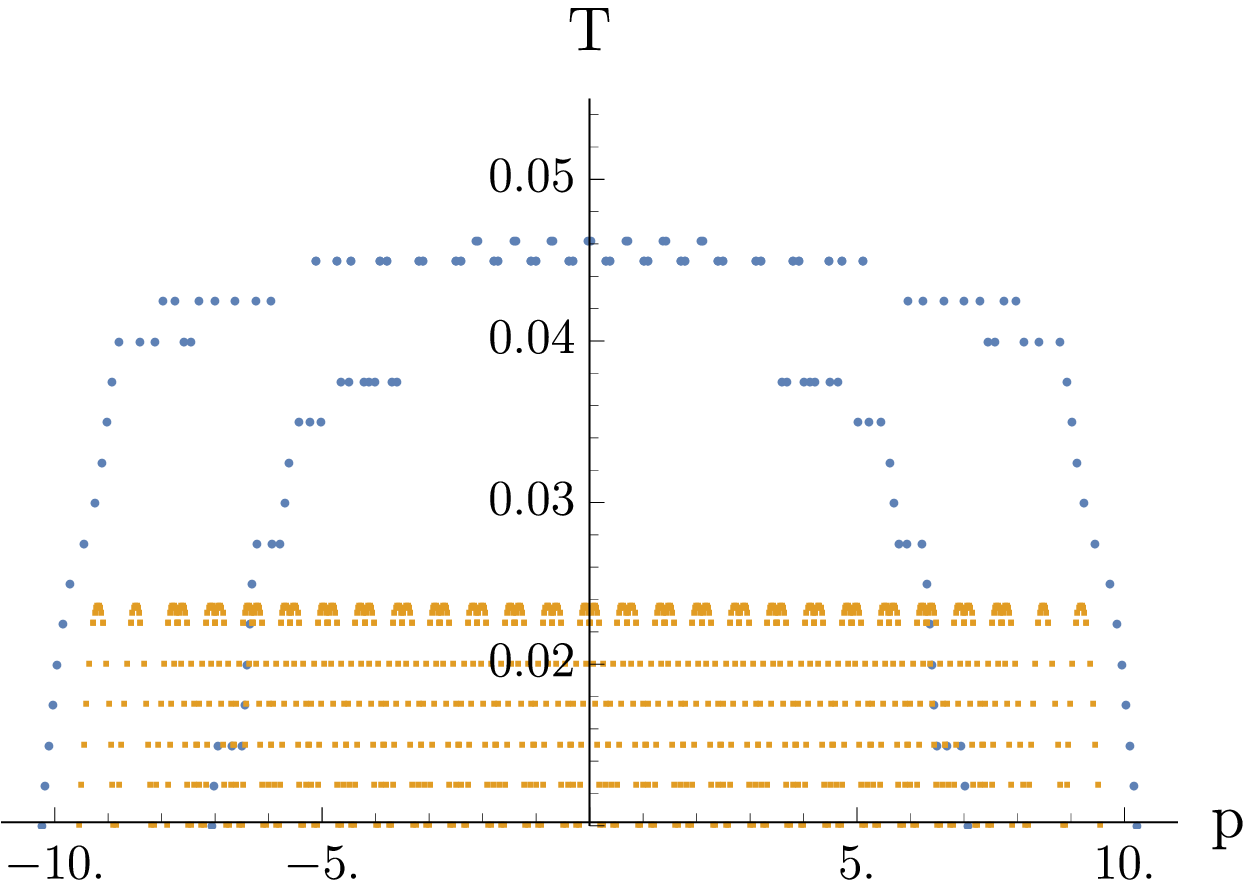}
\caption{\label{fig:HighPrecBell}Higher resolution $N_x = 27$}
\end{subfigure}
\caption{\label{fig:PrecBell} Sample results for the eigenvalue search for the backgrounds with $k=0.7$ and two amplitudes: $A=0$ (yellow dots) and $A=2.$ (blue dots).}
\end{figure}

By looking closely at the definition of the Bloch wave-functions \eqref{eq:marginal modes}, \eqref{eq:mm} we observe that if the eigenvalue $p^*$ has an eigenfunction $\delta \Phi^{p^*}(x)$, then there exists eigenvlue $p^* + nk$ with eigenfunction $ e^{-i n k x} \delta \Phi^{p^*}(x)$. At large $n \gtrsim N_x/2$ this eigenfunction becomes highly oscillatory and can not be accounted for by the discretization grid with $N_x$ nodes in $x$-direction. This is why one can numerically obtain only a finite number of eigenvalues. On Fig.\,\ref{fig:PrecBell} we show the samples of the actual output of \texttt{Eigenvalues} which we obtain after discretising the linear problem on the coarser $N_x=17$ and finer $N_x = 27$ grids (only real eigenvalues are shown). Correspondingly, $~10$ and $~20$ Brillouin zones are seen in the scope. The zero amplitude result (yellow dots) reveals indeed a number of copies of the RN instability curves, shifted by $n k$, $n \in (-N_x/2, N_x/2)$.

When we consider finite amplitudes $A$ the situation becomes more complicated since the modes with different momenta become coupled, as in \eqref{eq:matrix_coupling}. At larger values of amplitude the couplings between further separated modes come into play and the resulting band is a result of hybridisation of the several modes. At this point it becomes crucial for our study to use the discretization grid which would accommodate enough modes to allow for such hybridisation. We see indeed on Fig.\,\ref{fig:LowPrecBell}, that at finite amplitude (blue points) the band is not formed even the eigenvalues found in the first Brillouin zone feel the presence of the boundaries in momentum space. When we increase the resolution, Fig.\,\ref{fig:HighPrecBell}, more modes are taken into account and the band is robustly formed in first three Brillouin zones. We use this empirical criterion (evident formation of the band) in order to estimate the necessary size of the calculation grid, when collecting the data shown on Fig.\,\ref{fig:TPplot}.

\bibliographystyle{JHEP-2}
\bibliography{inhom_stripes_lattice}

\end{document}